\documentclass[aip,amsmath,amssymb,reprint]{revtex4-1}

\usepackage{graphicx}
\usepackage{dcolumn}
\usepackage{bm}
\usepackage[utf8]{inputenc}
\usepackage[T1]{fontenc}
\usepackage{mathptmx}
\usepackage{etoolbox}
\usepackage{multirow}
\usepackage{color}
\usepackage{xcolor}
\usepackage{amsmath}
\usepackage{graphicx}
\usepackage{amssymb}
\usepackage{textcomp}
\usepackage{float}
\usepackage{natbib}
\usepackage{ragged2e}
\usepackage{color,soul}
\usepackage{array}
\usepackage{dcolumn}
\usepackage{url}
\usepackage{multirow}
\usepackage{appendix}
\usepackage[normalem]{ulem}
\usepackage{soul}
\usepackage{caption}
\usepackage{subcaption}

\makeatletter
\def\@email#1#2{%
 \endgroup
 \patchcmd{\titleblock@produce}
  {\frontmatter@RRAPformat}
  {\frontmatter@RRAPformat{\produce@RRAP{*#1\href{mailto:#2}{#2}}}\frontmatter@RRAPformat}
  {}{}
}%
\makeatother
\begin{document}

\preprint{AIP/123-QED}

\title[]{Modeling photoassociative spectra of ultracold NaK+K
}

\author{Baraa Shammout*}
\affiliation{Institut für Quantenoptik, Leibniz Universität Hannover, 30167 Hannover, Germany}
\email{shammout@iqo.uni-hannover.de}

\author{Leon Karpa}
\affiliation{Institut für Quantenoptik, Leibniz Universität Hannover, 30167 Hannover, Germany}

\author{Silke Ospelkaus}
\affiliation{Institut für Quantenoptik, Leibniz Universität Hannover, 30167 Hannover, Germany}

\author{Eberhard Tiemann}
\affiliation{Institut für Quantenoptik, Leibniz Universität Hannover, 30167 Hannover, Germany}

\author{Olivier Dulieu*}
\affiliation{Universit$\acute{\text e}$ Paris-Saclay, CNRS, Laboratoire Aim$\acute{\text e}$ Cotton, Orsay, 91400, France}
\email{olivier.dulieu@universite-paris-saclay.fr}

\date{\today}

\begin{abstract}

A model for photoassociation of ultracold atoms and molecules is presented, and applied to the case of $^{39}$K and $^{23}$Na$^{39}$K bosonic particles. The model relies on the assumption that photoaossociation is dominated by long-range atom-molecule interactions, well outside the chemical bond region. The frequency of the photoassociation laser is chosen close to a bound-bound rovibronic transition from the $X^1\Sigma^+$ ground state toward the metastable $b^3\Pi$ lowest excited state of $^{23}$Na$^{39}$K, allowing to neglect any other excitation which could hinder the photoassociation detection. The energy level structure of the long-range $^{39}$K$\cdots$$^{23}$Na$^{39}$K excited super-dimer is computed in the space-fixed frame by solving coupled-channel equations, involving the coupling between the $^{23}$Na$^{39}$K internal rotation with the mechanical rotation of the super-dimer complex. A quite rich structure is obtained, and the corresponding photoassociation rates are presented. Other possible photossociation transitions are discussed in the context of the proposed model.

\end{abstract}

\maketitle

\section{Introduction}
\label{sec:intro}

Photoassociation (PA) of particles A and B (which could be either atoms or molecules) in a dilute gas is a light-induced process leading to the creation of a molecular complex AB by absorption of a photon with energy $h\nu$: A+B+$h\nu \rightarrow$ AB$^*$, where $h$ is the Planck constant and $\nu$ is the photon's frequency. In most cases the AB complex is left in an excited state (thus the star symbol) due to the energy deposited by the photon. PA is a powerful way to induce unimolecular reactions, being the inverse process of photodissociation, both pertaining to the so-called half-collision, as elegantly discussed in Ref.\onlinecite{nesbitt2012}. A sufficiently monochromatic light source can indeed populate a well defined quantum state of the AB$^*$ complex. But a limitation immediately occurs at room temperatures: the broad width of the kinetic energy distribution of the particles, covering many bound levels of the complex, drastically hinders the possibility to prepare a well-defined quantum state of AB$^*$ \cite{marvet1995,ban1999}.

The ground-breaking development of laser cooling of atoms for more than forty years immediately appeared as an exquisite opportunity to use PA as a tool to study ultracold gases composed of alkali-metal atoms\cite{thorsheim1987,miller1993,lett1993}. The kinetic energy distribution of ultracold atoms is now narrower than most energy level spacings of the cold atom pair, which can efficiently absorb a photon to populate a molecular bound level (bound-free transition) in a quasi-resonant way, similar to a bound-bound transition. PA soon became an important high-resolution molecular spectroscopy technique: it allowed the population of weakly-bound molecular levels \cite{stwalley1999,jones2006} with large spatial extension, as the atoms in an ultracold gas spend most of their time at distances much larger than the usual chemical bonds. PA spectroscopy thus advantageously complemented the few attempts to reach such levels via conventional molecular spectroscopy \cite{elbs1999}. Moreover, PA was the first approach to create samples of ultracold ground-state molecules \cite{fioretti1998,carr2009,dulieu2009}, well before the method based on magnetoassociation, which also leads to the formation of ultracold molecules in selected individual quantum states \cite{ni2008,carr2009,dulieu2009}.

The opportunity to study atom-molecule collisions in the ultracold regime is a natural extension to atom-atom studies. Several experiments clearly observed losses in trapped molecular samples induced by the presence of atoms  \cite{staanum2006,zahzam2006,hudson2008,ospelkaus2010a,knoop2010,deiglmayr2011}, which are presumably induced by atom-molecule scattering resonances \cite{yang2019,gregory2021,wang2021,nichols2022}. There is a vast literature about the modeling of atom-molecule collisions in the cold or ultracold regime. Focusing our interest on collisions involving ultracold alkali-metal atoms, which are quite heavy, open-shell, and with a strong electron-nuclear spin coupling, theorists suggested that such atom-molecule systems could be governed by a large number of scattering resonances, due to the large amount of available rovibrational states of the diatom \cite{mayle2012}, requiring their statistical treatment. Such a highly resonant character has indeed been observed \cite{yang2019,voges2020,wang2021,su2022,son2022}, and triatomic NaK$_2$ molecules have even been stabilized via such resonances \cite{yang2022a,yang2022b}. The model of Ref. \onlinecite{mayle2012} relies on the separation of the dynamics at large distances treated via quantum scattering theory, from the dynamics at short distances treated via a formalism involving random matrix theory. Alternate models attempting to represent interactions and scattering resonances in a refined way have been proposed \cite{soldan2002,quemener2005,levinsen2009,morita2018,frye2021}.

The purpose of the present paper is to explore under which conditions PA could be an efficient approach (as suggested in an earlier paper \cite{perez-rios2015}) to study mixtures of ultracold atoms and diatomic molecules, as a natural extension of atom-atom PA, and to create stable ultracold triatomic molecules. PA could also provide information about collisional dynamics, by populating well-defined quantum levels just below an excited dissociation threshold A+B$^*$. From a classical point of view, these weakly-bound levels are excited at large interparticle distances, so that the motion starts inward with almost vanishing local kinetic energy. In this context, an ultracold collision starts with very small kinetic energy at infinity. In both situations, \textit{i.e.}, colliding free particles, or weakly-bound particles, the system is sensitive to the strong short-range ''chemical'' interactions in the same manner. These interactions could thus be studied with PA as a function of the ''initial energy'', \textit{i.e.} the binding energies of the weakly-bound levels, which could span a larger range than the ones that can be reached in a full collision. In this latter case, the dependence of the dynamics as  function of the  initial kinetic energy cannot usually be studied, being constrained by the experimental setup. Instead, one uses magnetic Feshbach resonances (MFRs) \cite{stwalley1976,tiesinga1993}, occurring when an external magnetic field shifts the energy of weakly-bound levels of the particle pair with respect to the energy of the initial state, to reach a resonance with the initial kinetic energy.

\begin{figure}[htb]
    \centering
    \includegraphics[scale=0.7]{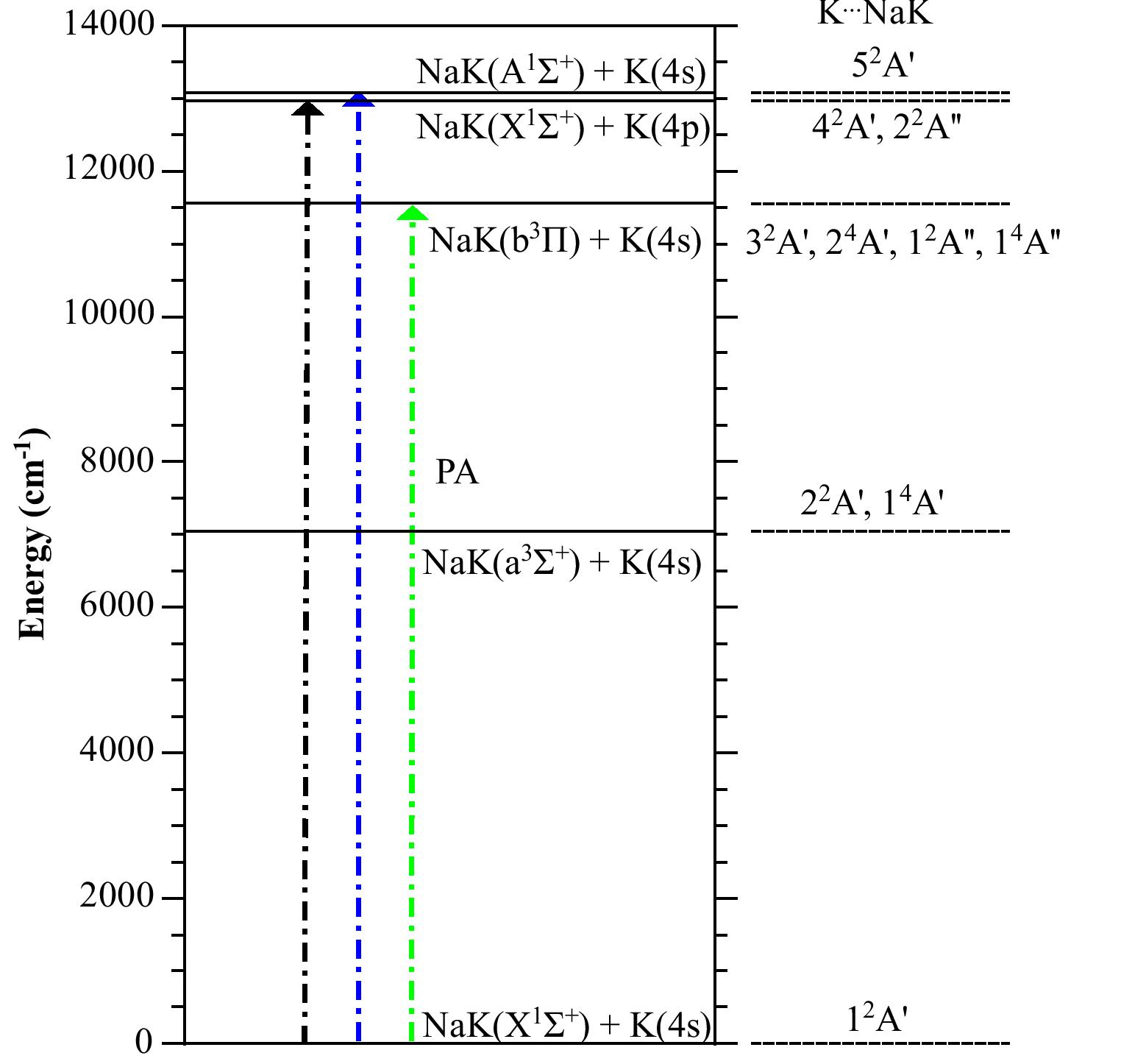}
    \caption{Simplified diagram of the energy levels of a K($4s$) or K($4p$) atom, combined with a NaK molecule in the lowest electronic states $X^1\Sigma^+$, $a^3\Sigma^+$, $b^3\Pi$,  and $A^1\Sigma^+$. The NaK energies are taken at the bottom of the $X^1\Sigma^+$ potential energy curve (PEC), $r_e(X)=6.6$~a.u., very close to the one of the $b^3\Pi$ PEC, but very different than the one of the other PECs (see Table \ref{tab:constants}). The vertical arrows depict possible vertical transitions from the $X^1\Sigma^+$ state. In green, the proposed atom-molecule photoassociation (PA) transition: it is clearly distinct from the transition which would allow for PA of K atoms (in black). An alternate PA transition (in blue) could concern the NaK($A^1\Sigma^+$)+K($4s$) limit, but its energy at the chosen distance is close to the excitation energy of the K atom. The K$\cdots$NaK electronic states defined in the $C_s$ symmetry group representation (Section \ref{ssec:longrangePECS}) and correlated to the limits above are listed on the right. A more comprehensive correlation diagram is reported in Appendix \ref{app:NaK_PEC}.}
    \label{fig:excitation_diagram}
\end{figure}

The main step of our theoretical approach, as in Refs. \onlinecite{perez-rios2015,elkamshishy2022}, only considers explicitly the long-range interactions between the atom and the molecule, while the short-range interactions are modeled via a boundary condition at a distance where the electronic exchange interaction is still negligible, around the so-called LeRoy radius \cite{leroy1970}. This hypothesis of the dominant long-range interaction in ultracold atom-molecule Feshbach resonances has been recently devised for ground-state K-NaK collisions \cite{wang2021}. To exemplify our approach, we consider ultracold $^{23}$Na$^{39}$K ground-state molecules immersed in a cloud of ultracold ground-state $^{39}$K atoms, which has been recently experimentally realized \cite{voges2022}. Figure \ref{fig:excitation_diagram} shows some of the lowest energy levels of the K-NaK pair. In contrast to previous work \cite{perez-rios2015,elkamshishy2022}, where PA is studied using  laser frequencies slightly detuned to the red of an atomic transition (black arrow in Fig. \ref{fig:excitation_diagram}), we choose a PA laser frequency close to a molecular transition (green arrow in Fig. \ref{fig:excitation_diagram}): we thus address a ''clean'' spectral range outside of the one of the atom-atom PA. 

In the present case, the PA laser frequency $\nu$ is chosen close to the molecular transition frequency between the lowest rovibrational level $v_X=0, j_X=0$ of a $^{23}$Na$^{39}$K molecule in its electronic ground state $X^1\Sigma^+$ (which can be experimentally prepared in suitable ensembles with high phase space density \cite{voges2020}) and the lowest rovibrational level $v_b=0, j_b=1$ of the $\Omega=0^+$ component of lowest excited electronic state $b^3\Pi$ (where $\Omega$ refers to the projection of the $^{23}$Na$^{39}$K total electronic angular momentum on the diatomic molecular axis). Thus, the search for PA signals will not be hindered by the presence of NaK transitions, as the $b^3\Pi$ state is the lowest of all excited electronic states. At large atom-molecule distances, the transition electric dipole moment (TEDM) of $^{23}$Na$^{39}$K determines the strength of the PA transitions: the $b^3\Pi$ state is weakly coupled by spin-orbit interaction to the neighboring $A^1\Sigma^+$ excited state, making this transition dipole-allowed. A PA scheme relying on the $A^1\Sigma^+$ state could be more difficult to identify because of overlap with the diatomic signal.

In Section \ref{sec:longrangestructure} we present our approach to derive the long-range potential energy curves (PECs) of the K-NaK complex based on advanced quantum chemistry methods. The results are collected in Section \ref{ssec:longrangePECS}, and the asymptotic model for the calculation of bound levels of the NaK-K complex in Section \ref{ssec:pa_spectrum}. Finally, the corresponding PA rates are shown in Section \ref{ssec:pa_rate} in the context of future experimental investigations.

In the following, we will omit the atomic masses and simply invoke K and NaK instead of $^{39}$K and $^{23}$Na$^{39}$K. Unless otherwise stated, distances will be given in atomic units 1~a.u.~$\equiv a_0=0.0529177210903(80)$~nm, with $a_0$ the Bohr radius, and energies in cm$^{-1}$, a convenient unit for spectroscopy, or in atomic units (a.u.) with 1~a.u.$\equiv$ 1~Hartree$=2\times 109737.31568160(21)$~cm$^{-1}$. The electric dipole moment expressed in a.u gives 1~a.u.=2.54175~D. These conversions are based on tabulated fundamental constants \cite{NIST_CUU}.

\section{Methods}
\label{sec:longrangestructure}

\begin{figure}[ht]
    \centering
    \includegraphics[scale=0.8]{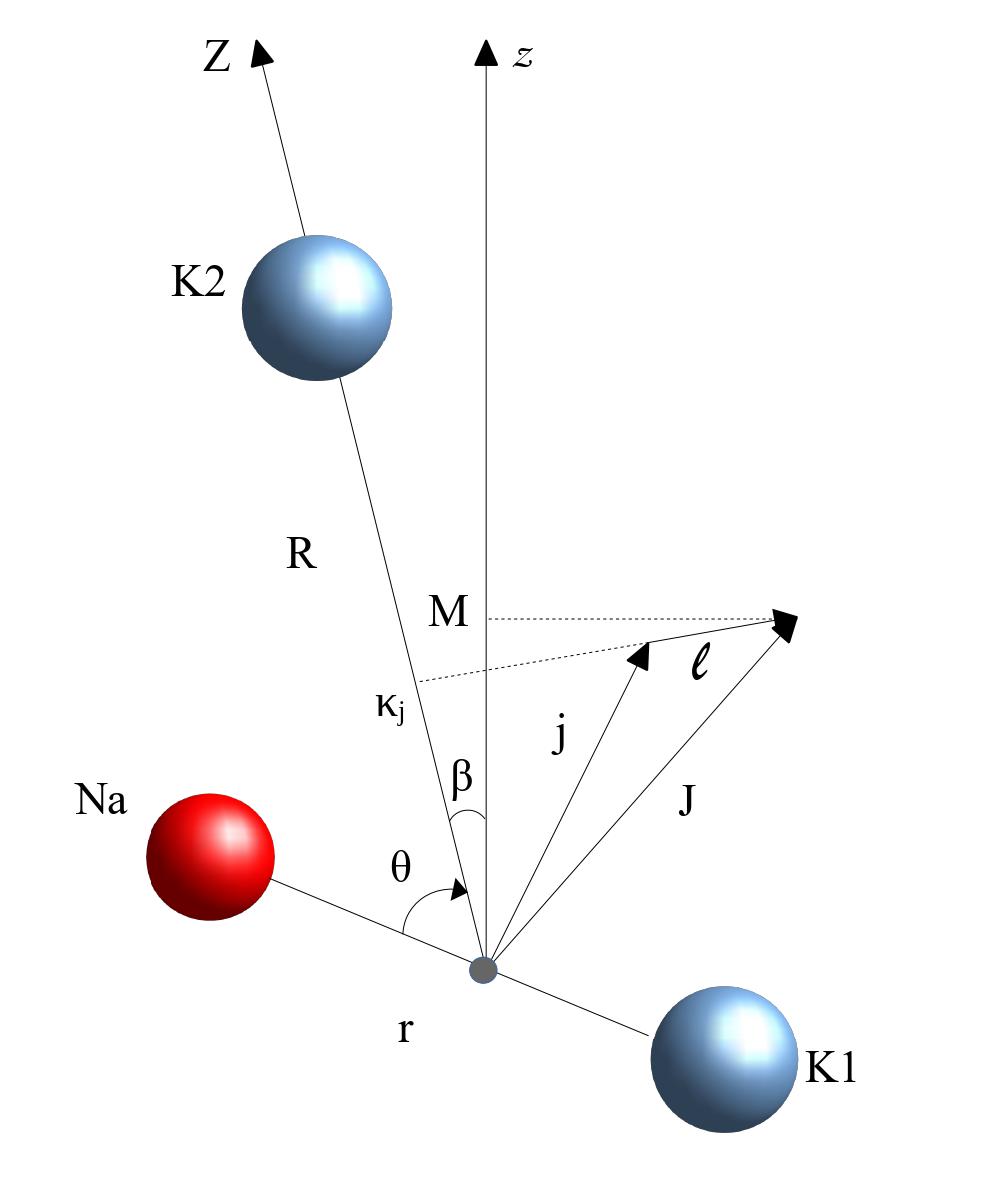}
    \caption{Chosen coordinates for the triatomic system NaK-K. The two K atoms are distinguished, which is consistent with a long-range approach. The Jacobi coordinates defined in the body-fixed (BF) frame ($XYZ$) are $R$, $r$ and $\theta$.  The space-fixed (SF) frame ($xyz$) is characterized by the Euler angle $\beta$ between the BF and SF axis $Z$ and $z$. The total (electronic+rotation) angular momentum $\vec{j}$ of NaK (with projection $\kappa_j$ on the BF $Z$) is coupled to the mechanical rotation $\vec{\ell}$ of the NaK-K pair. In the absence of external field, the resulting total angular momentum $\vec{J}=\vec{j}+\vec{\ell}$ is conserved, with a projection $\kappa_j$ (resp. $M$) on the BF $Z$ (resp. SF $z$) axis.}
    \label{fig:coordinates}
\end{figure}

We present in this Section the chosen approach for the electronic structure calculations of NaK and K-NaK. The K-NaK complex is described in Jacobi coordinates (Fig.~\ref{fig:coordinates}), $R$ being the distance between the potassium atom K (noted K2) and the center of mass of the diatom NaK (with the K atom noted K1), $r$ the bond length of NaK, and $\theta$ the angle between the vector $\vec{R}$ pointing toward K2, and the diatomic axis pointing from Na to K1. Thus $\theta=0$  and $\theta$=180$^{\circ}$ correspond to the linear configurations K-NaK and NaK-K, respectively. We always assume that $R >>r$. At such large distance $R$, the K-NaK interaction is much smaller than the energy separation between the $\Omega=0$ and $\Omega=1$ spin-orbit components of the $b^3\Pi$ state: $\Omega=0$ is taken as a conserved quantum number, so that it is not necessary to consider the projection of the total electronic angular momentum of the diatom on the $Z$ Jacobi axis. This approximation is sometimes identified as the super dimer model (see for instance Ref.\onlinecite{ayouz2011}, treating a similar situation). 

We compute the electronic structure with the MOLPRO software package \cite{werner2012,werner2020}. The triatomic complex NaK-K is modeled as a three-valence-electron system, where the electrons of the atomic ion cores are replaced by large effective-core relativistic pseudopotentials (ECP) referenced as ECP18SDF and ECP10SDF \cite{fuentealba1982}, for K$^+$ and Na$^+$, respectively. We use the valence basis sets associated with the ECPs in their uncontracted form as implemented in MOLPRO. We added $spdf$ diffuse functions with exponents reported in Tab.\ref{tab:Basis set}. In order to account for electronic correlations between the core and the valence electrons, we employ core polarization potentials (CPPs) \cite{fuentealba1982}, parametrized by the electric dipole polarizabilities $\alpha$ of Na and K, and cut-off radii $\rho$ for each species (Table \ref{tab:Basis set}).

\begin{table}[]
\begin{ruledtabular}
\begin{tabular}{l|llll|ll}
\multirow{2}{*}& \multicolumn{4}{l|}{Exponents} & 
\multicolumn{2}{l}{CPP parameters in a.u.} \\ 
\cline{2-7} 
& \multicolumn{1}{l|}
{$s$\footnote[1]{Adopted from Ref.~\onlinecite{zuchowski2010}}}   & \multicolumn{1}{l|}{$p$\footnotemark[1]} & 
\multicolumn{1}{l|}{$d$}                 & 
$f$                                      & 
\multicolumn{1}{l|}{$\alpha$ }  & 
$\rho$ \\ \hline
Na& \multicolumn{1}{l|}{0.009202} & \multicolumn{1}{l|}{0.005306}& \multicolumn{1}{l|}{0.3, 0.07}& 0.09 & \multicolumn{1}{l|}{0.9947}& 1.27   \\ 
K& \multicolumn{1}{l|}{0.009433} & \multicolumn{1}{l|}{0.004358}& \multicolumn{1}{l|}{0.38, 0.04} & 0.04 & \multicolumn{1}{l|}{5.354}& 1.86   \\ 
\end{tabular}
\end{ruledtabular}
\caption{For each species Na and K, exponents for the $spdf$ diffuse functions completing the basis set implemented in MOLPRO, and dipole polarizabilites $\alpha$ and cut-off radii $\rho$ defining the core polarization potentials (CPPs) \cite{fuentealba1982}.}
\label{tab:Basis set}
\end{table}

The potential energy curves of NaK, and the long-range potential energy surfaces (PESs) of NaK-K complex are calculated using the multiconfiguration reference internally contracted configuration interaction (MRCI) method \cite{werner1988} with Pople correction. The initial guess for orbitals is generated by the multi-configuration space-consistent field (MCSCF) method \cite{werner1985}. As we are dealing with long-range PESs, a good test of the appropriateness of the used basis sets at the MRCI level is given by comparing our results to other determinations of the PECs for the $X^1\Sigma^+$ and $b^3\Pi$ states of NaK. This is exemplified in Table~\ref{tab:constants} where the main spectroscopic constants of NaK are compared to the experimental ones, showing a satisfactory agreement of better than 1\%. In Appendix \ref{app:NaK_PEC}, we provide a direct comparison of the PECs, which demonstrates a satisfactory agreement over the entire PECs.
 
\begin{table}
\begin{ruledtabular}
\begin{tabular}{l c l l l}
 & & This work & Exp. & Ref. \\ \hline
\multirow{2}{*}{$X^1\Sigma^+$ } & $r_e$ (a.u.)  & 6.58 & 6.612217(3) &\onlinecite{yamada1992}  \\ 
& $\omega _e$ (cm$^{-1}$)&123.27&124.013(8) & \onlinecite{yamada1992}  \\ 
& $B_e$ (cm$^{-1}$)&  0.096 &  0.09522934(1) & \onlinecite{yamada1992}  \\
& $D_e$ (cm$^{-1}$)& 5259 &  5273.62(10) & \onlinecite{gerdes2008}  \\ 
\hline
\multirow{5}{*}{$b^3\Pi$}  & $r_e$ (a.u.) & 6.60& 6.62 &\onlinecite{ross1986a}  \\ 
& $\omega _e$ (cm$^{-1}$) & 120.21& 120.407(4) & \onlinecite{ross1986a} \\ 
& $B_e$ (cm$^{-1}$)& 0.096  & 0.09506(2)  &\onlinecite{ross1986a}   \\
& $T_e$ (cm$^{-1}$) & 11558& 11562.18  & \onlinecite{ross1986a}     \\
& $D_e$ (cm$^{-1}$)& 6666& 6697.9    &\onlinecite{ross1986a}     \\
\end{tabular}
\end{ruledtabular}
\caption{Computed spectroscopic constants for the $X^1\Sigma^+$ and $b^3\Pi$ states of $^{23}$Na$^{39}$K (this work), compared to various experimental data: equilibrium bond length $r_e$, potential well depth $D_e$, harmonic constant $\omega_e$, excitation energy $T_e$, and rotational constant $B_e$. For completeness, we found the minimum of the $A^1\Sigma^+$ PEC located at $r_e=7.93$~a.u..}
\label{tab:constants}
\end{table}

\section{Results}
\label{sec:results}

\subsection{Long-range potential energy surfaces of K $\cdots$ N\lowercase{a}K for the excited states $3^2A'$ and $1^2A''$ }
\label{ssec:longrangePECS}

At large distance, the weakly-bound K$\cdots$NaK complex (symbolized by the $\cdots$ symbol) has only the two symmetry operations: the identity operation, and the reflection through the plane containing the vectors $\vec{R}$ and $\vec{r}$. Hence we describe it in the framework of the C$_s$ point group. The expected lowest electronic trimer states are listed on the right of Fig.\ref{fig:coordinates}. The number in front of each symbol counts the states with equal symmetry from the bottom of the energy scale. We focus on the excited $3^2A'$ state that can be reached from the $1^2A'$ ground state by PA, and on the $1^2A''$ state for completeness, which correlate to the asymptote NaK($b^3\Pi$)+K($4s$) relevant for the chosen PA transition. Spin-orbit and hyperfine couplings are not introduced in the rest of the calculations. However, as stated in the introduction, the PA transition is allowed due to spin-orbit coupling between the $A^1\Sigma^+$ and $b^3\Pi$ states, giving rise to two states labeled as $0^+$. Thus for experimental implementation of the present model, the notation  $b^3\Pi$ should actually be understood as the component $b^3\Pi\,(0^+)$ of the triplet manifold. For simplicity, in the following, this component is labeled with the $b$ symbol only, while the $X^1\Sigma^+$, $a^3\Sigma^+$ and $A^1\Sigma^+$ are denoted with $X$, $a$, and $A$, respectively. The proposed experiment involves the lowest vibrational level $v_b=0$ of NaK($b$). The calculation mesh is defined in the following way: we vary the bond length $r$ over the extension of the vibrational wave function of this $v=0$ level, by taking 9 values between 5.804~a.u. and 7.436~a.u. in steps of 0.204~a.u.. A set of 19 values for $\theta$ between $0^{\circ}$ and $180^{\circ}$ with $10^{\circ}$ step size is adopted. We selected a variable grid step size $\delta R$ in $R$ adapted to the variation of the long-range PESs, with 29 values between 30~a.u. to 160~a.u. as follows: $\delta R=2$~a.u., 1~a.u., 5~a.u., 20~a.u., over the consecutive intervals [30$a_0$-40$a_0$], [40$a_0$-50$a_0$], [50$a_0$-100$a_0$], [100$a_0$-160$a_0$], respectively. In total, the three-dimensional long-range PESs are calculated on a mesh of $R\times r \times\theta=29\times9\times19=4959$ grid points.

In order to extract converged excited doublet states, we set the number of active orbitals to 6 (5 states in $A'$ and 1 in $A''$ irreducible representations). We perform state-averaged MCSCF of the lowest five $^2A'$ states using Configuration State Functions (CSF). Subsequently, two distinct multireference CI calculations are achieved for the four $^2A'$ lowest states, and for the lowest $^2A''$ state (including only the $2^2A''$ state in the internal CI).

\begin{figure}[h]
\includegraphics[scale=0.75]{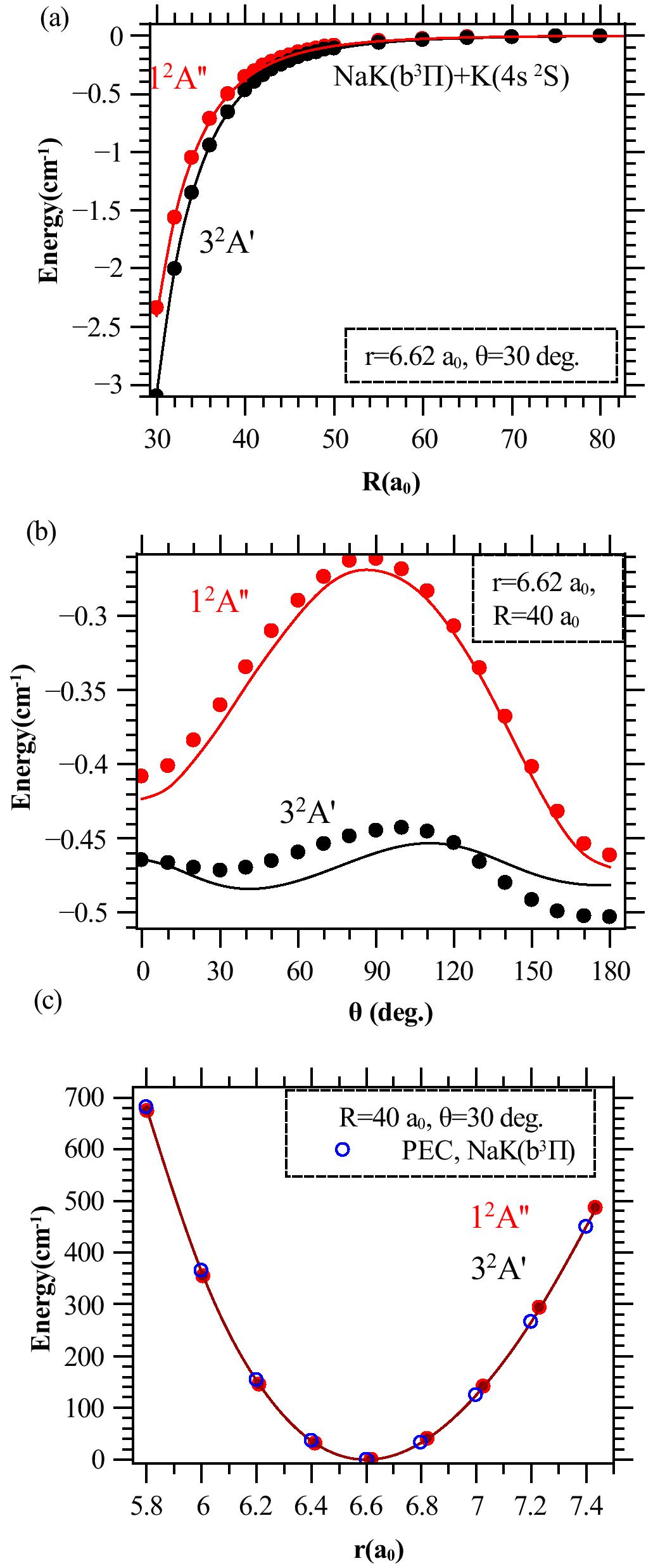}
\caption{One-dimensional cuts through the long-range PESs of the $3^2A'$(black circles) and $1^2A''$ (red circles) states of K$\cdots$NaK. The zero of energy is taken at the dissociation limit NaK($b^3\Pi$,$r=6.62$~a.u.)+K($4s$). Fits of the computed points according to Eq. \ref{eq:longrangepot} are displayed with solid lines. (a) At $r=6.62$~a.u.,$\theta=30^{\circ}$. (b) At $r=6.62$~a.u.,$R=40$~a.u. . (c) At $R=40$~a.u.,$\theta$=30$^{\circ}$, showing that the $3^2A'$ and $1^2A''$ PESs nicely match the NaK($b^3\Pi$) PEC (blue circles) over this $r$ interval, for such a large $R$ (see also Appendix \ref{app:NaK_PEC}).}
\label{fig:PES-cuts}
\end{figure}

In Fig.~\ref{fig:PES-cuts} we present three different cuts of the calculated long-range PES of the $3^2A'$ and $1^2A''$ states. The angular dependence of the PESs is exemplified in Fig.~\ref{fig:PES-cuts}b, at $r=r_e(b)=6.62$~a.u. and $R=40$~a.u.: the anisotropy of the $1^2A''$ PES is more pronounced than the one of the $3^2A'$ PES. It is worth noticing that the $3^2A'$ and $1^2A''$ PESs should be degenerate in the linear geometry: the observed differences are numerical errors reflecting the limited size of the chosen active space. They can be reduced by one order of magnitude by increasing the size of the active space to 7 states (5 states in $A'$ and 2 in $A''$ irreducible representations). However, such calculations are expensive, and as the qualitative behavior of the PESs is not significantly changed, we keep the present results. For completeness, we present the two-dimensional long-range PES (in $R$ and $\theta$) for these two states in Appendix \ref{app:NaK_PEC}.

For the fixed geometry $r=r_e(b)=6.62$~a.u. and $\theta=30^{\circ}$, the resulting cut of the $3^2A'$ PES is more attractive than the $1^2A''$ one for $R>30$~a.u. (Fig.~\ref{fig:PES-cuts}a). On the large energy scale of Fig.~\ref{fig:PES-cuts}c plotted for $R=40$~a.u. and $\theta=30^{\circ}$, both PES cuts look identical, approaching the $b$ PEC of NaK.

The calculated long-range PES can be fitted to the standard multipolar expansion expressed in atomic units of distance and energy,
\begin{equation}
    V(R,r,\theta )=-\frac{C_6(r,\theta )}{R^6}-\frac{C_8(r,\theta )}{R^8}+E_{\infty}(r),
\label{eq:longrangepot}
\end{equation}
where $E_{\infty}(r)$ is the $r$-dependent energy of the K$\cdots$NaK complex for $R\to \infty$, thus identical to $b$ PEC of NaK (Appendix \ref{app:NaK_PEC}). The conversion of the $C_n$ coefficients to other popular units is easy: $C_n \textrm{(in cm$^{-1}$\AA$^{-n}$)}  \equiv 2R_{\infty} a_0^n C_n \textrm{(in a.u.)}$. The dominant term controlled by the $C_6$ coefficient results from the Debye interaction \cite{debye1920a,debye1920b,lepers2018} of a permanent dipole inducing an instantaneous dipole on a non-polar particle. The parameters in Eq.\ref{eq:longrangepot} are obtained via a two-step procedure: (i) by fitting $V(R,r,\theta )$ between $R=60$~a.u. and $R=160$~a.u. to the dominant term $-\frac{C_6(r,\theta )}{R^6}+E(\infty)$, and (ii) keeping $C_6(r,\theta )$ and $E(\infty)$ fixed, and fitting $V(R,r,\theta )$ to Eq.\ref{eq:longrangepot} between $R=40$~a.u. and $R=160$~a.u.  to estimate $C_8(r,\theta )$. These fits are found accurate to better than 1\% at fixed $\theta$ (Fig.~\ref{fig:PES-cuts}a, c), while the fit of the angular dependence (Fig.~\ref{fig:PES-cuts}b) is slightly less satisfactory with a deviation of about 2\% to 4\%.

\begin{figure}
    \centering
    \includegraphics[scale=0.8]{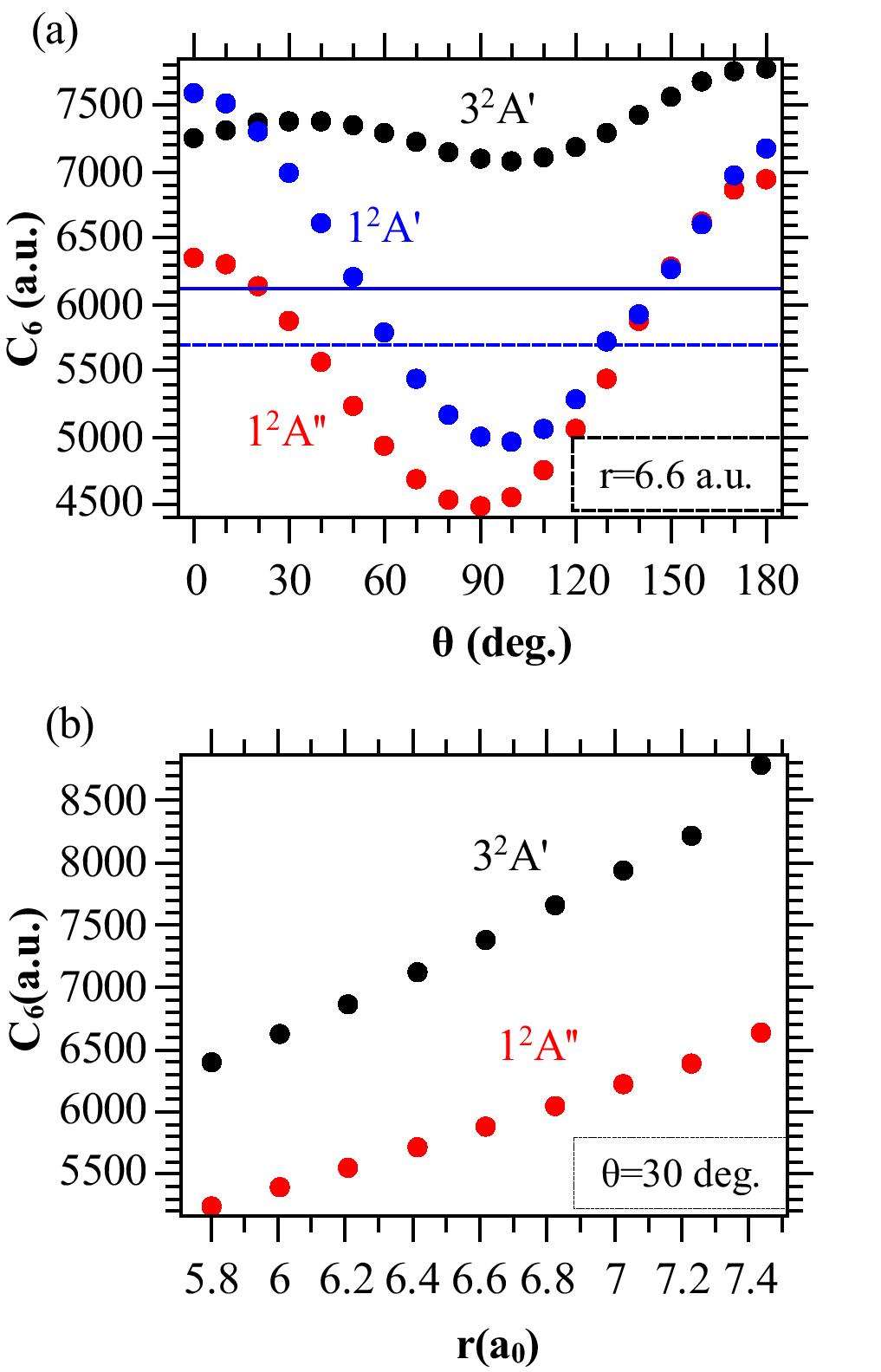}
    \caption{The $C_6$ coefficients of the long-range PES of the $1^2A'$ (blue circles) $3^2A'$ (black circles), $1^2A''$ (red circles) states of K$\cdots$NaK, (a) as functions of $\theta$ at $r=r_e(b^3\Pi)=6.62$~a.u. for $3^2A'$ and $1^2A''$ and at $r=r_e(X^1\Sigma^+)=6.61$~a.u. for $1^2A'$; (b) as functions of $r$ at $\theta=30^{\circ}$. In panel (a), the horizontal solid line gives the spherically-averaged value $C_6(\theta)$ for the ground state $1^2A'$ compared to the value of Ref.\onlinecite{zuchowski2013} (dashed line).}
    \label{fig:C6}
\end{figure}

Figure \ref{fig:C6} shows the fit results. As expected, the anisotropy of the PESs is reflected in the variation of $C_6(r=6.62\textrm{a.u.},\theta)$ (Fig. \ref{fig:C6}a), decreasing by about 30\% from $\theta=0^{\circ}$ to $\theta=90^{\circ}$ for the $1^2A''$ curve, and by about 5\% from $\theta=30^{\circ}$ to $\theta=90^{\circ}$ for the $3^2A'$ curve. Note that the numerical error invoked above induced by the limited size of the active space, results in an uncertainty of about 10\% on the value of $C_6(r,\theta )$. This value is consistent with the spherically-averaged value $\bar{C_6}=6103$~a.u., compared to the value $\bar{C_6}=5698$~a.u. from Ref. \onlinecite{zuchowski2013}. The results in Fig. \ref{fig:C6}b expresses the physics of the Debye interaction: the coefficient $C_6(r))$ (with $\theta=30^{\circ}$ in the figure) increases with $r$, as its dipole moment does in this region \cite{aymar2005}.

Equation \ref{eq:longrangepot} is useful for easily calculating the long range PESs at arbitrary values of $R$, $r$ and $\theta$, as this will be required for solving the Schrödinger equation for the atom-molecule relative motion in the next Section. 

\subsection{Weakly-bound energy levels of the K$\cdots$N\lowercase{a}K complex}
\label{ssec:pa_spectrum}


We treat the K$\cdots$NaK system in free space as an effective two-body problem (referred to as a super dimer model) in the space-fixed-frame $xyz$, assuming a total angular momentum $\vec{J}$. A ground-state K($4s\, ^2S$) atom, considered as structureless, approaching a diatomic molecule NaK in a given rovibrational level $(v_{\Bar{\Lambda}}=0,j)$ with energy $\epsilon_{vj}$ of an electronic state $\Bar{\Lambda}$, hinders the free rotation of the diatom, which generates anisotropy of the long-range K$\cdots$NaK interaction potential $V(R,r,\theta)$, thus coupling the NaK rotational levels. The relevant angular momenta $\vec{j}$, $\vec{\ell}$, $\vec{J}=\vec{j}+\vec{\ell}$, are defined in Fig. \ref{fig:coordinates}, with $|j-\ell| \leq J\leq j+\ell$. The corresponding operators will be denoted $\hat{J}$, $\hat{j}$,and $\hat{\ell}$.  In this Section we calculate the K$\cdots$NaK weakly-bound energy levels close to the dissociation limits K($4s$)+NaK($b(v_b=0,j_b)$) using a standard coupled-channel approach. 


The Schr\"odinger equation $\hat{H} \Psi=E\Psi$ for the K$\cdots$NaK system with eigenfunction $\Psi$ and energy $E$, involves the Hamiltonian

\begin{equation}
   \hat{H} =\hat{T}_R+\hat{T}_r+\hat{V},
    \label{eq:htrimer}
\end{equation}
with the kinetic energy operators associated with the $R$ and $r$ coordinates
\begin{eqnarray}
   \hat{T}_R =-\frac{\hbar^2}{2\mu R }\frac{\partial^2}{\partial R^2}R+\frac{\hbar^2\hat{\ell}^2}{2\mu R^2}, \\ \nonumber
   \hat{T}_r =-\frac{\hbar^2}{2\mu' r }\frac{\partial^2 }{\partial r^2}r+\frac{\hbar^2\hat{j}^2}{2\mu'r^2},
    \label{eq:Ttrimer}
\end{eqnarray}
and where $\mu=m(\textrm{K})m(\textrm{Na}\textrm{K})/[m(\textrm{K})+m(\textrm{Na}\textrm{K)}$] is the reduced mass of the complex K$\cdots$NaK and $\mu'$ is the reduced mass of NaK. The operator $\hat{V}$ corresponds to the interaction potential $V(R,r,\theta)$ between K and NaK, and includes the potential energy of the diatom, while the one of the isolated atom is disregarded as a fixed quantity. As in the previous section, we keep the zero of energies as the energy of the K$\cdots$NaK system for $R\to\infty$ with $r=r_e(\Bar{\Lambda})$, namely the location of the bottom of the PEC of the $\Bar{\Lambda}$ electronic state (see Fig. \ref{fig:PES-cuts}c, Appendix \ref{app:NaK_PEC}). In the following, the $\Bar{\Lambda}$ index will be removed for simplicity.

We first define the basis set in the BF frame for a given $J$ value with its projection $M$~\cite{Mvalue},
\begin{equation}
|J v j \kappa_{j} \rangle \equiv \frac{\chi_{ v j}(r)}{r}\sqrt{\frac{2J+1}{4 \pi}} D^{J*}_{M \kappa_j}(\alpha,\beta,\gamma) Y_{j,\kappa_j}(\theta,0).
\label{eq:BFbasis}
\end{equation}
Here $\chi_{v,j}(r)$ is the rovibrational wave function with eigenvalue $\epsilon_{vj}$, and $Y_{j,\kappa_j}(\theta,0)$ is a spherical harmonics associated with the rovibrational level $(v,j)$ of the $\Bar{\Lambda}$ electronic state of the isolated NaK molecule. The projection of $\vec{j}$ on the $Z$ BF axis is denoted with $\kappa_j$. The Wigner functions $D^{J*}_{M \kappa_j}(\alpha,\beta,\gamma)$ refer to the transformation between the SF and BF frames, and depend on the Euler angles $(\alpha,\beta,\gamma)$. The PES $V(R,r,\theta)$ can be recast in this basis set as a matrix with elements
\begin{equation}
V^{J}_{j \kappa_j,j' \kappa_j}(R)=\langle J v j' \kappa_j \left| V(R,r,\theta)-E_{\infty}(r)\right| J v j \kappa_j \rangle_{r,\theta},
\label{eq:PEC_BF}
\end{equation}
where the notation $\left< \left|\right|\right>_{r,\theta}$ denotes the integration over $r$ and $\theta$. The straightforward integration over Euler angles is performed, but not labeled, for simplicity, leading to diagonal terms in $\kappa_j$ only. The diagonal terms $V^{J}_{j \kappa_j,j \kappa_j}(R)$ represents the one-dimensional PECs of the super dimer. The off-diagonal elements $V^{J}_{j \kappa_j,j' \kappa_j}(R)$ hold for the couplings between NaK rotational levels induced by the anisotropy of the interaction potential.  As the energy range around the asymptote of K$\cdots$NaK addressed in the rest of the paper is very small compared to the energy spacing of the $\Bar{\Lambda}$ vibrational levels, we restrict the basis set to $v=0$, so that this index can be removed.

The BF basis set $|J j \kappa_j\rangle$ is transformed to the basis set $|J j \ell \rangle$ in the SF-frame according to \cite{launay1976,hutson1991,lara2015}
\begin{eqnarray}
|J j \ell \rangle =(2\ell+1)^{1/2} 
\times \sum_{\kappa}{}(-1)^{j-\ell-\kappa}\begin{pmatrix}
j & \ell & J\\
\kappa & 0 & -\kappa
\end{pmatrix}
\nonumber\\
\times |J j \kappa \rangle,
\label{eq:BF_SF}
\end{eqnarray}
where $\kappa_j \equiv \kappa$ as $\Vec{J}$ and $\Vec{j}$ have the same projection on the $Z$ axis, and the parenthesis refer to $3j$-Wigner symbols. The matrix elements $V^{J}_{j \kappa_j,j' \kappa_j}(R)$ are transformed to the SF frame,
\begin{eqnarray}
V^{J}_{j \ell,j' \ell'}(R)=(2\ell+1)^{1/2}(2\ell'+1)^{1/2}(-1)^{j-j'}(-1)^{\ell+\ell'} \nonumber\\
\times \sum_{\kappa}{}\begin{pmatrix}
j & \ell & J\\
\kappa & 0 & -\kappa
\end{pmatrix}
\begin{pmatrix}
j' & \ell' & J\\
\kappa & 0 & -\kappa
\end{pmatrix} V^{J}_{j \kappa, j'\kappa}(R).
\label{eq:PEC_SF}
\end{eqnarray}
In the SF frame, for a given $J$, the solution of the Schr\"odinger equation with an energy $E_n^J$ and total wave function $|J;E_n^J\rangle \equiv \Psi^J(R; E_n^J)$ can be expanded as
\begin{equation}
|J;E_n^J\rangle\equiv \Psi^J(R; E_n^J)=R^{-1}\sum_{j \ell }|J  j \ell \rangle\psi^J_{j\ell, n}(R),
    \label{eq:total_wavefunction}
\end{equation}
where the radial channel wave functions $\psi^J_{j\ell, n}(R)$ are solutions of the set of coupled equations
\begin{eqnarray}
\left[ \hat{T}_R+V^{J}_{j \ell,j \ell}(R)+\epsilon_{j}-E_n^J \right] \psi^{J}_{j \ell, n}(R) \nonumber\\
=-\sum_{j' \ell'}^{'}V^{J}_{j \ell,j' \ell'}(R) \psi^{J}_{j' \ell', n}(R)
\label{eq:Schrodinger_Eq}
\end{eqnarray}
\begin{equation}
 \langle J;E_n^J|J;E_n^J\rangle=1=\sum_{j\ell }\int  |\psi^{J}_{j\ell,n} (R)   |^{2}dR.
    \label{eq:norm}
\end{equation}
where we define the partial norm as
\begin{equation}
C^{J}_{j\ell,n} = \int |\psi^{J}_{j \ell,n} (R)  |^{2}dR.
\label{eq:partialnorm}
\end{equation}
and the rotational constant of the K$\cdots$NaK super dimer as
\begin{equation}
B_n^J= \langle J;E_n^J|1/(2\mu R^2)|J;E_n^J\rangle
    \label{eq:rotconstant}
\end{equation}

\begin{figure}
\includegraphics[scale=0.7]{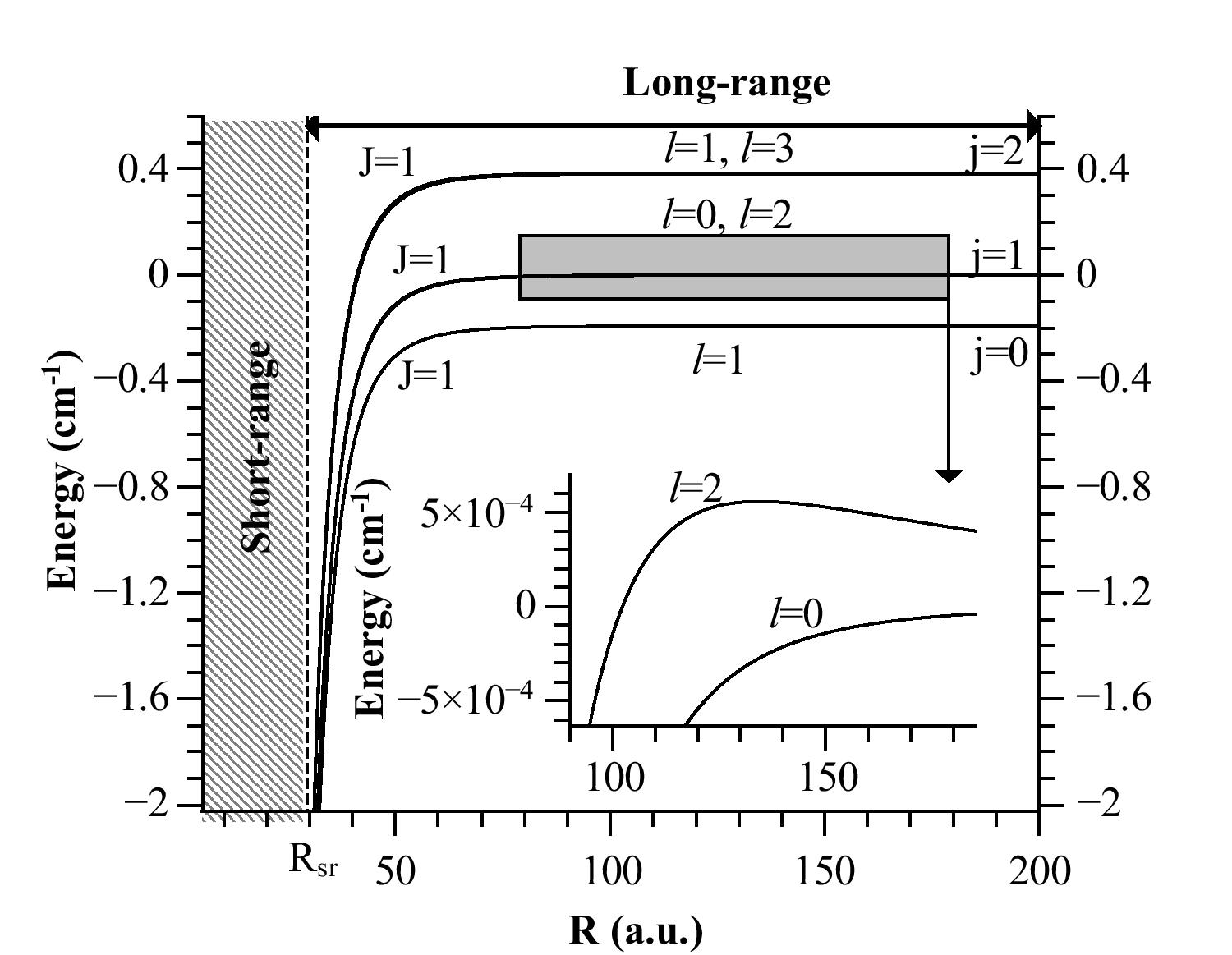}
    \caption{The effective long-range PECs of the NaK$\cdots$K system in the SF frame for $J=1$ as functions of $R$, $W^{1}_{j \ell, j \ell}(R)=V^{1}_{j \ell, j \ell}(R) +\epsilon_{j}+\ell(\ell+1)/(2\mu R^2)$, involved in the coupled-channel calculations of the weakly-bound states close to the NaK($b^3\Pi, v_b=0,j_b$)+K($4s$) limit (Eq. \ref{eq:Schrodinger_Eq}). The couplings between the channels are ignored for atom-molecule separation $R<R_{\textrm{SR}}$. The inset shows a magnified view of the grey box, with the contribution of the partial waves $\ell=0,2$.}
    \label{fig:Coupled_channels}
\end{figure}

The envisioned PA experiment starts from the $1^2A'$ state (Fig.\ref{fig:excitation_diagram}) with $J=0, j=0, \ell=0$, and thus positive parity. We focus on the weakly-bound energy levels of the $3^2A'$ and $1^2A''$ electronic states of the K$\cdots$NaK complex, with $J=1$ and negative parity $p=(-1)^{j+\ell}=-1$. After convergence checks, we limit the number of coupled equations to $N_{max}=5$ by including the five basis vectors $|J j l \rangle = |101 \rangle, |110 \rangle, |112 \rangle, |121 \rangle, |123 \rangle$. The five diagonal elements $W^{1}_{j \ell, j \ell}(R)=V^{1}_{j \ell, j \ell}(R) +\epsilon_{j}+\ell(\ell+1)/(2\mu R^2)$ in the Hamiltonian of Eq. \ref{eq:Schrodinger_Eq}, corresponding to these five channels, are displayed in Fig.\ref{fig:Coupled_channels}. This figure sets the range of energies where weakly-bound states could be calculated for this system in the framework of our long-range hypothesis, namely in a window no larger than 2~cm$^{-1}$.

The numerical solution of Eq.\ref{eq:Schrodinger_Eq} requires the definition of the interaction potentials for $R<R_{\textrm{sr}}$ (see Fig.\ref{fig:Coupled_channels}). We connect the diagonal terms $W^{J}_{j \ell, j \ell}(R)$ at $R_{\textrm{sr}}=30$~a.u. to short-range Lennard-Jones potentials (expressed in a.u.) of the form 
\begin{equation}
    V_{LJ}=D_{LJ}(C_{LJ}/R^6)[(C_{LJ}/R^6)-1]+E(\infty), 
\label{eq:LJ_pot}
\end{equation}
for which the coefficients are gathered in Appendix \ref{app:LJ_Parameters}. Still for numerical simplification, for $R<R_{\textrm{sr}}$, the off-diagonal terms $V^{J}_{j \ell, j' \ell'}(R)$ are kept equal to their values at $R_{\textrm{sr}}$. In Appendix \ref{app:prediss-res}, we illustrate that the eigenvalues of the system are insensitive to such a choice. 

We solve Eq.(\ref{eq:Schrodinger_Eq}) with the Mapped Fourier Grid Hamiltonian (MFGH) method, \cite{kokoouline1999} considering a grid in $R$ coordinate with 338 points, bounded by $R_{\textrm{min}}=5.6$~a.u. and $R_{\textrm{max}}=1000$~a.u.. The computed energies of the weakly-bound vibrational levels of the $3^2A'$ state, located below the NaK($b$, $v_b=0, j_b=0$)+K($4s$) limit, are presented in Table \ref{tab:Aprime} (listed above the horizontal line), while those of the $1^2A''$ state are displayed in Appendix \ref{app:levels_Adoubleprime}. The reported lowest bound level at -2.08~cm$^{-1}$ corresponds to an outer turning point of the PES at about 33~a.u., slightly outside the short-range region defined by $R_{\textrm{sr}}=30$~a.u., such that the eigenvalue may be already influenced by the chosen PEC matching with the short-range region. As it was used in previous studies using MFGH, the rotational constant $B_n$ reflects the spread in $R$ of the probability density, which decreases with the binding energy, namely as the radial wave function extends toward large distance. Its variation is not smooth as it depends on the channels composing each eigenstate. This composition is analyzed with the partial norm $C^{J}_{j\ell,n}$. Between $n=-15$ and -1, 11 levels are dominated by a single channel up to 90\% or more, revealing that the couplings between the channels are weak in most cases. This is consistent with the weak anisotropy observed for the $A'$ state (Fig. \ref{fig:PES-cuts}). In contrast, the more significant anisotropy of the $A''$ state is reflected in the strong channel mixing of the bound levels (Appendix \ref{app:levels_Adoubleprime}). The partial norm $C^{J}_{1 0,n}$ in Table \ref{tab:Aprime} is the only one to be considered for the present PA scheme, due to the dipole transition selection rules (see Section \ref{ssec:pa_rate}). Only five levels have a noticeable value of $C^{J}_{1 0,n}$ (larger than 0.2, labeled by a star), which thus could be expected to be detected in the experiment.

The MFGH method complemented with the stabilization method (see for instance Ref.\onlinecite{osseni2009}) allows for the localisation of the quasibound levels, or resonances, located between the NaK($b$, $v_b=0, j_b=0,1$)+K($4s$) limits, and listed below the horizontal line in Table \ref{tab:Aprime}. Their characterization and the possibility to detect them in the proposed PA scheme are discussed in Appendix \ref{app:prediss-res}.

\begin{table*}
\caption{Energies $E_{n}^J$ (with respect to the NaK($b^3\Pi$, $v=0, j=1$)+K($4s$) limit), rotational constants $B_{n}^J$ of the super dimer (Eq.\ref{eq:rotconstant}, multiplied by $10^{+3}$), and partial norms $C_{j \ell,n}^{J}$ (Eq.\ref{eq:partialnorm}) of weakly-bound vibrational levels (numbered from the uppermost one with a negative index, $n=-15$ to -1) of the $3^2A'$ ($J=1$) state, and located below the NaK($b^3\Pi$, $v=0, j=0$)+K($4s$) limit (at -0.19082~cm$^{-1}$ on this scale).  The stars label levels which could be seen in the proposed PA scheme. The same quantities are displayed for predissociating resonances ($n=$1 to 7) located between the NaK($b^3\Pi$, $v=0, j=0,1$)+K($4s$) limits (see Appendix \ref{app:prediss-res}).}
\begin{ruledtabular}
\begin{tabular}{ >{\raggedright\arraybackslash}p{0.03\linewidth} | >{\raggedright\arraybackslash}p{0.1\linewidth} | >{\raggedright\arraybackslash}p{0.1\linewidth}| >{\raggedright\arraybackslash}p{0.1\linewidth}||
>{\raggedright\arraybackslash}p{0.1\linewidth}||
>{\raggedright\arraybackslash}p{0.1\linewidth}|
>{\raggedright\arraybackslash}p{0.1\linewidth}|     
>{\raggedright\arraybackslash}p{0.1\linewidth}}
$n$&$E^1_{n}$ (cm$^{-1}$)& $10^{+3} B_{n}^1$(cm$^{-1})$ & $C_{01,n}^{1}$& $C_{10,n}^{1}$ & $C_{12,n}^{1}$& $C_{21,n}^{1}$& $C_{23,n}^{1}$  \\ \hline
-15  & -2.08302 & 3.9359 & \bf{0.85860} & 0.04450 & 0.09272 & 0.00161 & 0.00256 \\
-14  & -1.93277 & 3.9488 & 0.08529 & 0.00667 & \bf{0.88880} & 0.01050 & 0.00874 \\
-13 & -1.91967 & 3.9441 & 0.05326 & \bf{0.91628}* & 0.00132 & 0.01961 & 0.00953 \\
-12 & -1.64709 & 4.0032 & 0.00154 & 0.02128 & 0.00522 & \bf{0.96181} & 0.01015 \\
-11 & -1.63070 & 4.0099 & 0.00159 & 0.01144 & 0.01209 & 0.00620 & \bf{0.96869} \\
-10 & -1.22069 & 3.2778 & \bf{0.93890} & 0.02159 & 0.03743 & 0.00086 & 0.00122 \\
-9 & -1.06302 & 3.2486 & 0.06009 & 0.28494* & 0.65471 & 0.00001 & 0.00025 \\
-8 & -1.05143 & 3.2807 & 0.00028 & 0.67998*& 0.29990 & 0.01230 & 0.00754 \\
-7 & -0.74335 & 3.2866 & 0.00338 & 0.01264 & 0.00332 & \bf{0.97989} & 0.00077 \\
-6 & -0.72519& 3.2785 & 0.01062 & 0.00323 & 0.00691 & 0.00107 & \bf{0.97817} \\
-5 & -0.67031 & 2.5159 & \bf{0.94626} & 0.02846 & 0.01150 & 0.00375 & 0.01003 \\
-4 & -0.50220 & 2.5616 & 0.02348 & 0.72385* & 0.25078 & 0.00173 & 0.00017 \\
-3 & -0.48620 & 2.5563 & 0.00342 & 0.25827* & 0.72796 & 0.00526 & 0.00509 \\
-2 & -0.35828 & 1.8134 & \bf{0.98749} & 0.00426 & 0.00733 & 0.00040 & 0.00053 \\
-1 & -0.22396 & 1.0952 & \bf{0.97948} & 0.00865 & 0.01037 & 0.00080 & 0.00070 \\ \hline \hline
1 & -0.18263  & 0.5123 & 0.74792 & 0.22281 & 0.01269 & 0.01498 & 0.00161 \\
2 & -0.17464 & 0.8462 & 0.57744 & 0.00769 & 0.37116 & 0.02853 & 0.01519 \\
3 & -0.14644 & 2.1803 & 0.13784 & 0.06911 & 0.03322 & 0.74452 & 0.01531 \\
4 & -0.13015  & 2.2584 & 0.12186 & 0.02058 & 0.03386 & 0.00206 & 0.82164 \\
5 & -0.03704 & 0.9015 & 0.20546 & 0.70224 & 0.09101 & 0.00108 & 0.00021 \\
6 & -0.03150  & 0.6516 & 0.43635 & 0.10115 & 0.45990 & 0.00091 & 0.00168 \\
7 & -0.00033  & 0.3247 & 0.07172 & 0.92341 & 0.00467 & 0.00012 & 0.00007  
\end{tabular}
\end{ruledtabular}
\label{tab:Aprime} 
\end{table*}

\subsection{Photoassociation rate of K and N\lowercase{a}K}
\label{ssec:pa_rate}

We consider the initial state $i$ to be a ground state K atom and a ground state NaK molecule absorbing a photon while approaching each other at large separation $R$ to create a weakly-bound level of the K$\cdots$N\lowercase{a}K electronically-excited complex in a final state $f$ (Eq. \ref{eq:total_wavefunction}). Under typical experimental conditions \cite{voges2022}, the number of ultracold NaK molecules in the ultracold sample is much  smaller than the number of atoms, so that we define the number density of the minority particles as $n_{\textrm{NaK}}$. In this photoassociation (PA) process, the photon energy $h \nu_{\textrm{PA}} = h c/ \lambda_{\textrm{PA}}$ is assumed to be slightly smaller than the one of a NaK electronic transition which will be specified below. Following the above Sections, the diatom actually absorbs the photon while it is perturbed by the atom, so that the transition electric dipole moment (TEDM) $\vec{d}_{\textrm{NaK}}$ of NaK characterizes the strength of the chosen PA transition. Therefore the PA rate $R_{if}$, \textit{i.e.} the number of K$\cdots$NaK electronically-excited complexes per unit time and per diatomic molecule, can be calculated in a similar way as for atom-atom PA \cite{pillet1997,cote1998,perez-rios2015,elkamshishy2022}. Its expression in SI units (s$^{-1}$), for an ultracold atom-molecule sample at temperature $T$ exposed to a PA laser with intensity $I_{\textrm{PA}}$, and Boltzmann-averaged, is 
\begin{eqnarray}
R_{if}(T)& = &\frac{4 \pi^2 h^2}{c \epsilon_0} \frac{1}{(2\pi\mu k_BT)^{3/2}}  \\ \nonumber
&\times &n_{\textrm{NaK}} I_{\textrm{PA}} (d^q_{\textrm{NaK}})^2 |S_{if}(E_r)|^2 e^{-E_r/k_BT},
\label{eq:PArate}
\end{eqnarray}
involving the Planck constant $h$, the speed of light $c$, the Boltzmann constant $k_B$, and the vacuum permittivity $\epsilon_0$. As above, $\mu$ refers to the reduced mass of the K$\cdots$NaK complex. The energy $E_r$ is the atom-molecule collision energy which satisfies the resonance condition for the PA transition. The $q$ cartesian component of {$\vec{d}_{\textrm{NaK}}$} characterizes the active transition in the NaK diatom, namely the $X \to A$ electronic transition in the present case. The squared integral
\begin{equation}
|S_{if}(E_r)|^2 = \left[\int_{R_{\textrm{sr}}}^{R_{\textrm{max}}}  \psi^J_{10,n}(R)\xi_i(R, E_r)dR\right]^2,
\label{eq:overlap}
\end{equation}
expresses the spatial overlap (restricted to the long-range region relevant for the present study)  between the continuum radial wave function $\xi_i(R, E_r)$ of the the K$\cdots$NaK complex of the entrance channel, and the relevant radial component $\psi^J_{10,n}(R)$ of the total wave function $\Psi_f(R,E_n)$ of Eq. \ref{eq:total_wavefunction}. 

It is useful to derive the PA rate $K_{if}(T)$ normalized to the photon flux $I_{\textrm{PA}}  \lambda_{\textrm{PA}}/(hc)$ and to the molecular density $n_{\textrm{NaK}}$, in SI units (m$^5$) as 
\begin{equation}
K_{if}(T) = \frac{4 \pi^2 h^3}{\epsilon_0} \frac{1}{(2\pi\mu k_BT)^{3/2}} \frac{1}{\lambda_{\textrm{PA}}} d_{\textrm{NaK}}^2 |S_{if}(E_r)|^2 e^{-E_r/k_BT},
\label{eq:PArate_norm}
\end{equation}
to discuss the PA efficiency independently of particular experimental conditions.

\begin{figure}
    \centering
\includegraphics[scale=0.60]{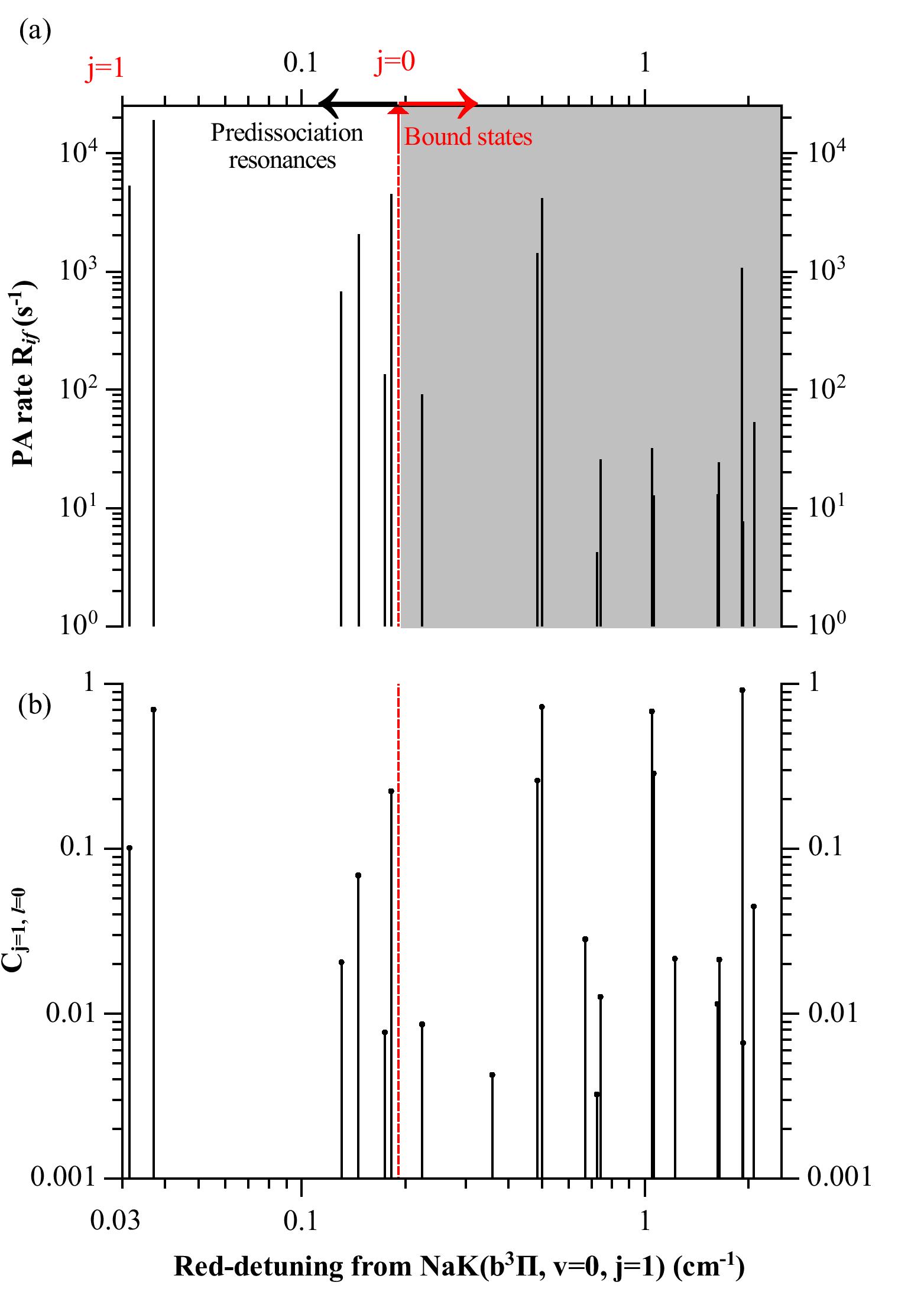}
    \caption{(a) Photoassociation rate $R_{if}(T=200\,\textrm{nK})$ (Eq.\ref{eq:PArate}) of NaK+K as a function of the red detuning (in cm$^{-1}$, displayed as positive values due to the log scale) of the PA laser with respect to the transition $X,v_X=0, j_X=0 \to b,v_b=0,j_b$=1, for a PA laser intensity $I_{PA}=100$~W.cm$^{-2}$ and a density of minority particles NaK of $n_{\textrm{NaK}}=10^{11}$~cm$^{-3}$, representative of the experimental conditions \cite{voges2022}. The energy position of the $b,v_b=0,j_b=0$ level (vertical red arrow) on this scale is 0.1908~cm$^{-1}$. We computed bound levels (above this value, grey area) and predissociating resonances (below this value, white area). (b) The partial norm $C_{10,n}^{1}$ from Table \ref{tab:Aprime}, at the same energy scale.}
    \label{fig:PA-rate}
\end{figure}

In the initial state $i$ of PA we assume that the K($4s$) atom collides in the $s$-wave regime ($\ell=0$) with a NaK molecule in the $(v_X=0,j_X=0)$ lowest rovibrational level of its electronic ground state $X$. It is thus represented by a single scattering channel with $J=0$, which is identical in the BF and SF frames. Its PES results from the previous section, with a long-range $C_6^X(\theta, r=6.6\textrm{ a.u.})$ coefficient displayed in Fig.\ref{fig:C6}a. As we disregard the short-range interactions above, we model the entrance channel with a single PEC of Lennard-Jones type in its original form \cite{lennard-jones1931} $V_X(R)=(\bar{C}_6^X)^2/(4 \epsilon_d R^{12})-\bar{C}_6^X/R^{6}$. The isotropic coefficient $\bar{C_6^X}=6119$~a.u. is obtained by spherically-averaging $C_6^X(\theta)$, and shows reasonable agreement with the value $\bar{C}_6^X=5698$~a.u. of Ref. \onlinecite{zuchowski2013} obtained from dynamic polarizability calculations. The energy $\epsilon_d=2201$~cm$^{-1}$ is of similar magnitude as the well depth of the NaK$_2$ electronic ground state with respect to the energy of K($4s$)+NaK($X$,$r=6.6$~a.u.) that we computed along the lines of Section \ref{sec:longrangestructure}. The resulting energy-normalized radial wave function $\xi_i(R, E_r)$ is computed with the standard Numerov integration method between $R_{\textrm{min}}$ and $R_{\textrm{max}}$. The transition dipole moment is taken from the calculations of Ref.~\onlinecite{vexiau2017}, revealing that the $v_b=0$ level of the $b$ electronic state contains a fraction of $\zeta=1.5 \times 10^{-4}$ of the $A$ electronic state, so that $(d^q_{\textrm{NaK}})^2= \zeta \times (d^q_{\textrm{NaK}})^2(X \to A)$, where $d^q_{\textrm{NaK}}(X \to A)$ is the TEDM between the $X$ NaK ground state and its $A$ excited state coupled to the $b$ state by spin-orbit interaction, as previously quoted (Eq. \ref{eq:PArate}). As the bottom of the  $X$ and $b$ PECs have very similar shape and an almost equal equilibrium distance (Table \ref{tab:constants}), the relevant vibrational wave functions of the dimer perfectly overlap, so that we chose $d^q_{\textrm{NaK}}(X \to A)=3.818$~a.u. at $r=6.6$~a.u. \cite{aymar2007}.
 
Due to the dipolar transition selection rules ($\Delta J=\pm 1,\Delta j=\pm 1,\Delta \ell=0$), only the channel wave function component associated with $|J=1,j=1,\ell=0 \rangle$ contributes to the squared integral (Eq. \ref{eq:overlap}). The results for the PA rate (in s$^{-1})$ for conditions relevant to the proposed experiment, are displayed in Fig. \ref{fig:PA-rate}(a). Assuming an average collision energy $E=k_B \times 200$~nK between K and NaK, with a molecular density $n_{mol}=10^{11}$~cm$^{-3}$ and a PA laser intensity $I_{PA}=100$~W/cm$^2$, the expected photoassociation rate reaches up to several thousands events per second for three weakly-bound levels. Figure \ref{fig:PA-rate}(b) reports the values of the partial norm $C_{10,n}^{1}$ over the same energy range. The difference of patterns between the two panels, in particular the change of the relative amplitude of the results for the highest bars in panel (a), compared to panel (b), illustrates the constructive (for the $n=-13, -4$ levels) or destructive (for the $n=-8$ level) interference between the initial and final radial wave functions (See Appendix \ref{app:radial-wf}).

The normalized PA rate $K_{if}$ is at most on the order of $10^{-30}$~cm$^5$, thus about five orders of magnitude smaller than the results obtained in cases where the PA transition is tuned to the atomic resonance \cite{perez-rios2015,elkamshishy2022}. It is directly related to the ratio between the molecular TEDM, $d^q_{\textrm{NaK}}(X \to b)= \zeta^{1/2} d^q_{\textrm{NaK}}(X \to A) \approx 0.046$~a.u., and the atomic one, the latter being two orders of magnitude larger than the molecular one. The expected density of PA resonances is lower in our case, as the long-range atom-diatom PECs vary as $R^{-6}$, compared to the $R^{-5}$ behavior of the quadrupole-dipole interaction induced by the quadrupole moment of the atomic $^2P$ state \cite{perez-rios2015,elkamshishy2022}.

\section{Discussion and conclusion}
\label{sec:discussion}

We argue below that the proposed PA scheme is expected to induce a detectable signal under the conditions of our reference experiment \cite{voges2022}. The main difference of our approach in comparison to the one of Ref. \onlinecite{perez-rios2015,elkamshishy2022} is that we choose to tune the PA laser frequency close to a molecular transition of NaK, instead of an atomic $^2S \to ^2P$ transition, to get rid of any PA line associated with the formation of K$_2$ dimers. Moreover the chosen NaK transition, $X,v_X=0,j_X=0 \to b^3\Pi(0^+), v_b=0, j_b=1$, reaches the vicinity of the lowest possible excited rovibrational level of NaK, such that there is only a single open dissociation channel nearby for the photoassociated trimer, namely, K($4s\,^2S$)+NaK($b^3\Pi(0^+), v_b=0, j_b=0$). Due to our hypothesis of dominant long-range interactions, it is unlikely that the hyperfine structure would play a significant role in the PA process. Indeed, both the initial and final states involve NaK molecules with a projection $\Omega=0$ of the total electronic angular momentum, inducing hyperfine splittings with a magnitude of a few tens of kHz \cite{aldegunde2017}.

The PA signal will result from the loss of the weakly-bound photoassociated trimers from the optical trap, following their subsequent spontaneous emission within typically a few tens of ns. This loss signal will emerge from a background signal free from other diatomic resonant processes, as stated above. The predicted PA rate of about 1000~s$^{-1}$ is larger than the typical loss rate of the ground-state NaK molecules from the trap, 10~s$^{-1}$, and thus fast enough to induce a detectable signal. If the photoassociated atom-molecule bound level is dominated by long-range interactions, as assumed in the present model, narrow PA lines should then be recorded. These lines may be slightly broadened by the possible predissociation toward  the neighboring K($4s\,^2S$)+NaK($b^3\Pi(0^+), v_b=0, j_b=0$) channel, as it has been already reported for potassium-potassium PA \cite{wang1998}. It could also happen that the PA laser addresses weakly-bound levels which are actually strongly dominated by short-range interaction, which thus would not appear anymore as narrow isolated lines, but instead as a broad profile containing many closely spaced resonances. They would contribute to an increase of the molecule decay rate in an apparently non-resonant manner. In total, we find that PA signals of atom-molecule system should be observable starting from a molecular transition and will have a structure that is not too dense and allows for a meaningful interpretation..

An alternate PA scheme is to tune the PA laser close to a dipole-allowed  transition of NaK, such as $X(v=0,j=0) \to A(v=0,j=1)$, or $X(v=0,j=0) \to B^1\Pi(v=0,j=0)$ (labeled as $B$ in the following). However, while the corresponding TEDM $d^q_{\textrm{NaK}}(X \to A)(r)$ and $d^q_{\textrm{NaK}}(X \to B)(r)$ are sizable \cite{aymar2007}, the minima of the $A$ and $B$ PECs are not aligned with the one of the $X$ PEC, leading to small overlap of the corresponding diatomic vibrational wave functions, and thus reducing the transition dipole moment (see Eq. \ref{eq:overlap}). These values would presumably lead to PA rates of  magnitude comparable to the ones of the present PA scheme. A full modeling of these options will be treated in a forthcoming work. Moreover, numerous predissociation channels would then be opened for the photoassociated weakly-bound trimers, which could broaden the PA lines in a noticeable way. 

Instead of looking for a trap-loss signal to probe PA, one option, inspired by the experimental demonstration of Ref. \onlinecite{hu2019}, could be to use a UV laser pulse to ionize the photoassociated weakly-bound trimers, resulting in easily detectable diatomic or triatomic ions. As in the pioneering experiment of PA of cesium atoms \cite{fioretti1998}, a sufficiently long time delay between the PA laser and the ionizing laser pulse could also probe the formation of ultracold ground state NaK$_2$ trimers, created after the spontaneous decay of the photoassociated weakly-bound trimers. We already computed the full PESs of the NaK$_2$ trimers following the method described in Section \ref{sec:longrangestructure}, which will be reported in an upcoming paper.
 \section*{Acknowledgements}
We gratefully acknowledge Dr Romain Vexiau for sharing his calculations on NaK transition dipole moments, and Charbel Karam for his support in using MFGH code. B. S., L. K and S. O. gratefully acknowledge financial support from Germany’s Excellence Strategy – EXC-2123 QuantumFrontiers – 390837967, the Deutsche Forschungsgemeinschaft (DFG) through CRC 1227 (DQ- mat), project A03, and the European Research Council through the ERC Consolidator Grant 101045075 TRITRAMO.

\appendix

\section{Additional results on electronic structure calculations }
\label{app:NaK_PEC}

\begin{figure*}
    \centering
    \includegraphics[scale=0.6]{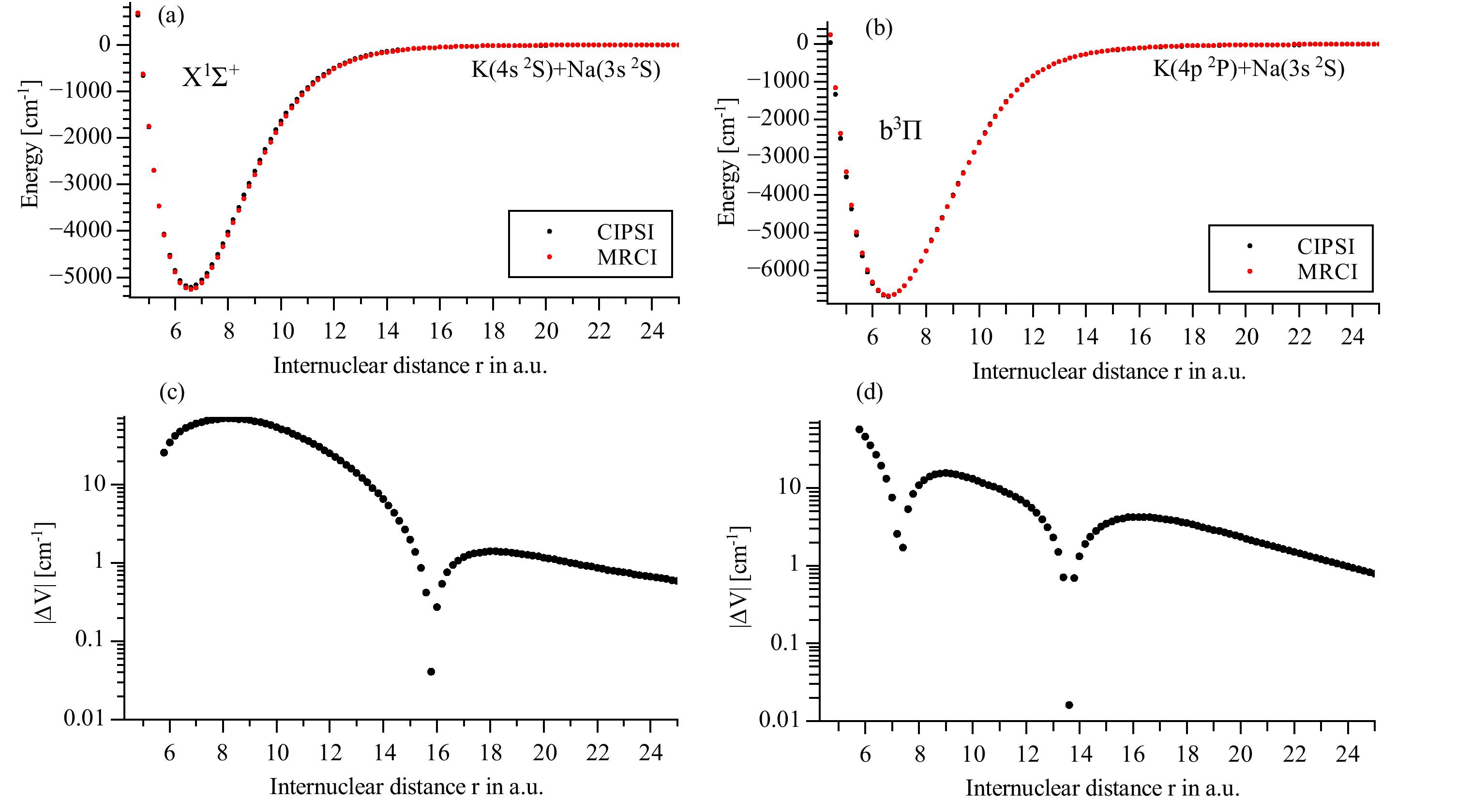}
    \caption{Computed PECs for (a) the  $X^1\Sigma^+$ ground state and (b) the $b^3\Pi$ excited state of NaK from the present work (MRCI, red dots), and from Ref.\onlinecite{aymar2007,vexiau2017} (CIPSI, black dots), and the energy difference $\Delta V =V_{\textrm{MRCI}}-V_{\textrm{CIPSI}}$ for (c) $X^1\Sigma^+$, and (d) $b^3\Pi$, respectively. The zero of energy is taken at the corresponding asymptotes Na($3s$)+K($4s$) and Na($3s$)+K($4p$), respectively.}
    \label{fig:PECs-NaK}
\end{figure*}

In Fig.~\ref{fig:PECs-NaK} we compare our calculated PECs for the $X$ and $b$ electronic states of NaK with the results obtained with the CIPSI approach (Configuration interaction by perturbation of a multiconfiguration wave function selected iteratively), with large ECP and additional atomic-state-dependent CPPs \cite{aymar2007,vexiau2017}. The overall agreement is very satisfactory, as it was anticipated from Table~\ref{tab:constants} of the main text.

\begin{figure}[h]
    \centering
    \includegraphics[scale=0.11]{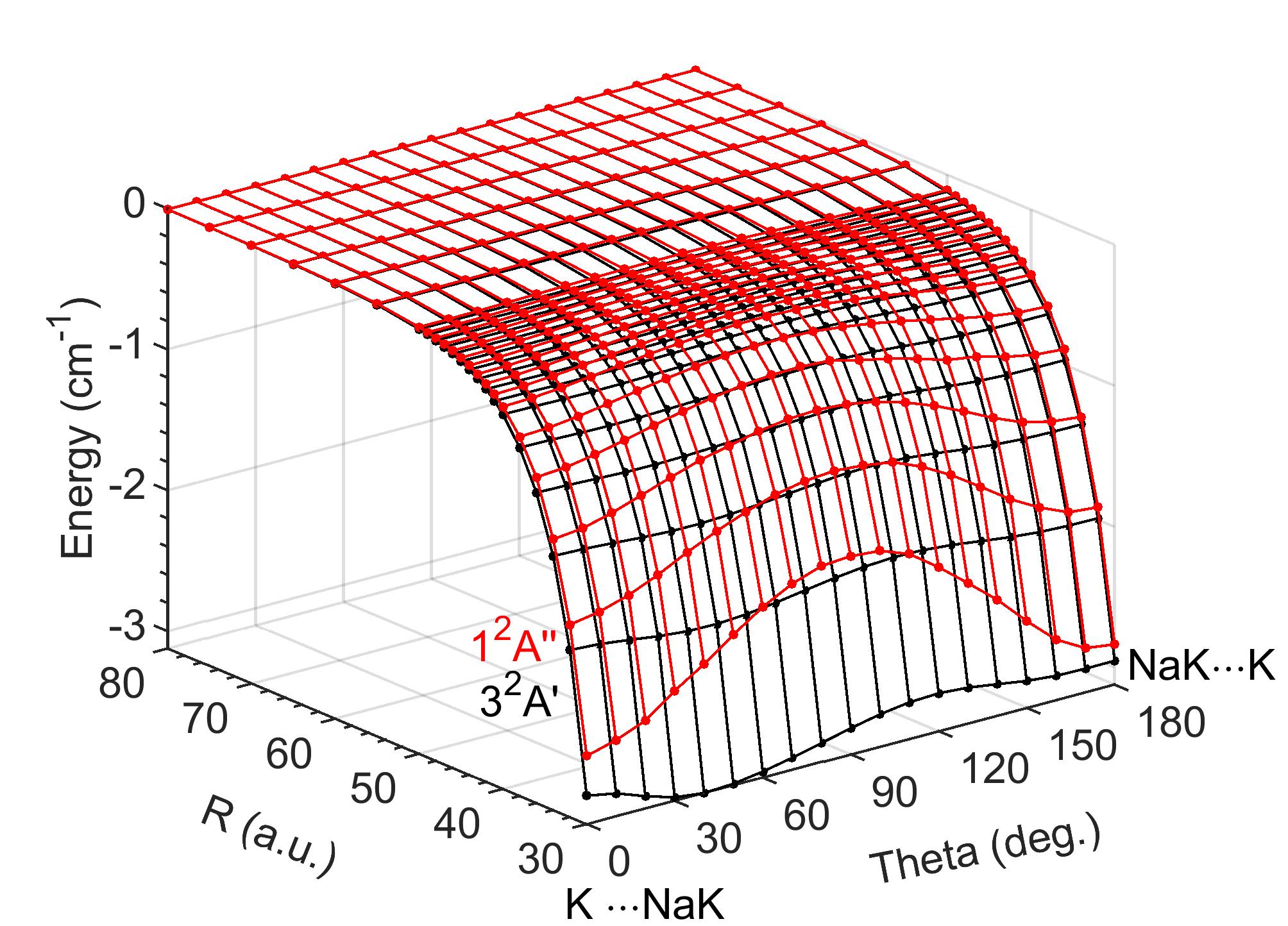}
    \caption{Computed long-range PESs for the $3^2A'$ and $1^2A''$ excited states at $r=r_e(b)=6.62$~a.u..}
    \label{fig:2D-PESs_NaK-K}
\end{figure}

Figure \ref{fig:2D-PESs_NaK-K} displays the two-dimensional PESs of the $3^2A'$ (in black)  and $1^2A''$ (in red) at $r=r_e(b)=6.62$~a.u., where the isopotential lines reflect the different anisotropy of the two PESs at large distances,  as noticed in Fig.~\ref{fig:PES-cuts} of the main text. A more detailed presentation of the full PESs will be presented in a forthcoming paper.

\begin{figure}[h]
    \centering
    \includegraphics[scale=0.5]{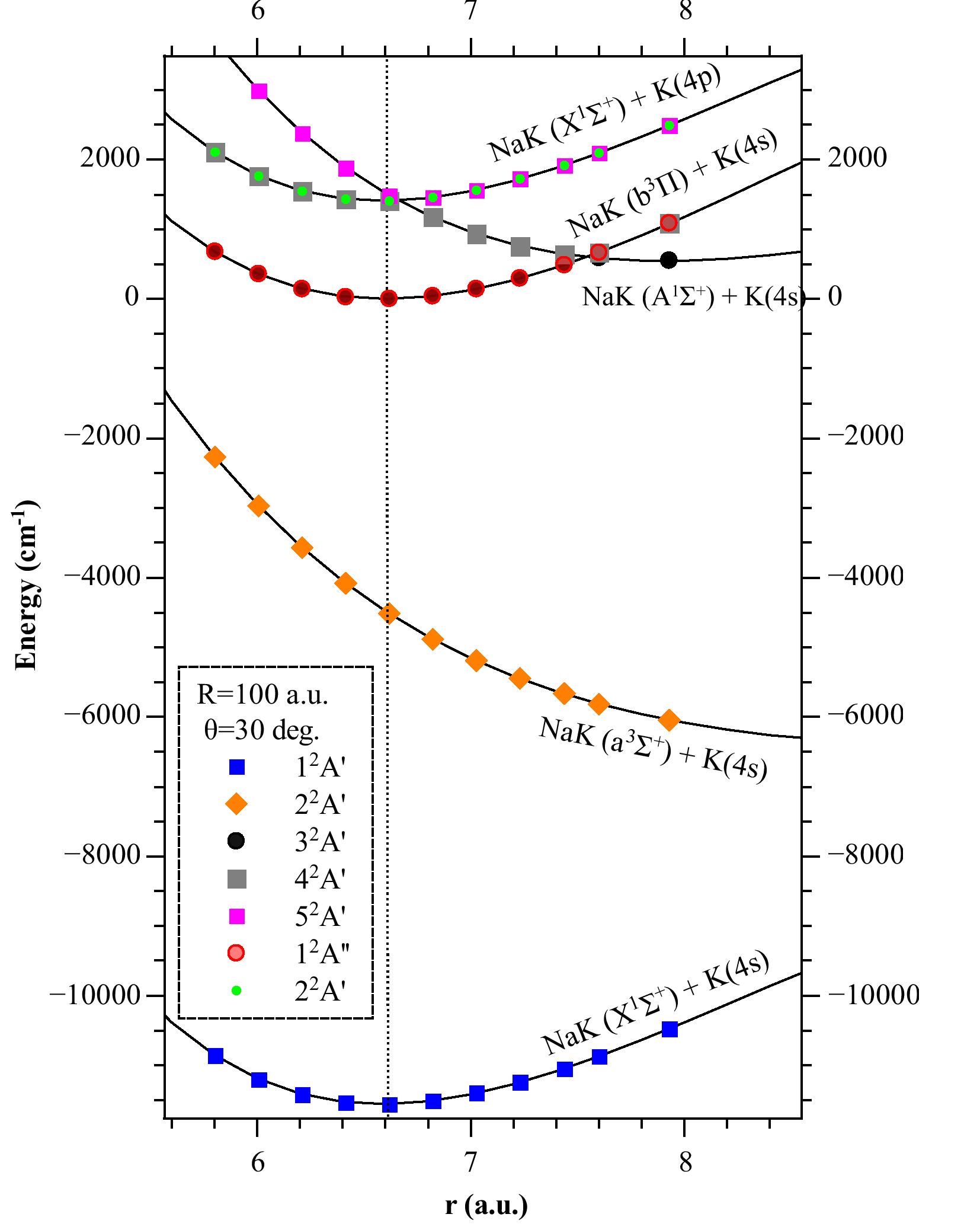}
    \caption{Correlation diagram between the electronic states of the K$\cdots$NaK trimer and those of the separated K atom and NaK dimer. The symbols depict one-dimensional cuts of the computed long-range PESs of K$\cdots$NaK, in $r$ coordinate in the interval [5.80~a.u.-7.93~a.u.], and $R=100$~a.u., $\theta=30$°, for the five lowest $A'$ states and the two lowest $A''$ states. The zero of energies is fixed at the energy of NaK($b$,$r=6.62$~a.u.)+K($4s$).  The solid lines hold for the computed PECs of NaK in the lowest electronic states $X^1\Sigma^+$, $a^3\Sigma^+$, $b^3\Pi$, $A^1\Sigma^+$ (accompanied by a free K($4s$) atom at infinity), as well as of the $X^1\Sigma^+$ PEC shifted by the excitation energy of K($4p$). The vertical line marks the position of the minimum of the $X$ and $b$ PECs of NaK.}
    \label{fig:correlation_diagram}
\end{figure}

In Fig.~\ref{fig:PES-cuts}c of the main text, we already invoked, that our calculations of the $3^2A'$ PES nicely converge for large $R$ toward the $b$ PEC of NaK, over a restricted range of $r$ values. We actually explored further this check of the accuracy of the present MRCI computations by calculating the PESs of K$\cdots$NaK at $R=100$~a.u., where the NaK PECs should be reproduced independently of the angle $\theta$. We extended our computations by increasing the number of active orbitals to 7 (\textit{i. e.} 5 (2) states for the $A'$ ($A''$) irreducible representations, respectively), and we performed the state-averaged MCSCF of the lowest seven $^2A'$ states and the lowest three $^2A''$ states. We achieved a good convergence of the five lowest $^2A'$ PESs, and of the two lowest $^2A''$ PESs, toward the lowest NaK PECs. Figure \ref{fig:correlation_diagram} thus presents an extension of both Figs. \ref{fig:excitation_diagram} and \ref{fig:PES-cuts}c of the main text, resulting in a correlation diagram. Due to the avoided crossing between the $3^2A'$ and $4^2A'$ states at $r=7.55$~a.u. resulting from the known crossing between the $b$ and $A$ PECs, the correlation of the $3^2A'$ switches from the $b$ state (as assumed in the present paper) to the $A$ state beyond this distance. As anticipated in Fig.~\ref{fig:excitation_diagram} of the main text, the $4^2A'$ and $5^2A'$ PESs undergo an avoided crossing at $r=6.67$~a.u., close to the equilibrium distance of the $X$ and $b$ PECs of NaK. It corresponds to the crossing between the PA channel where K is excited, and the one where NaK is excited in the $A$ state.

\section{Lennard-Jones potentials at short-range}
\label{app:LJ_Parameters}
Table~\ref{tab:short-range} lists the parameters of the Lennard-Jones potentials used to extend the diagonal elements of the effective potentials $W^{J}_{j \ell, j \ell}(R)=V^{J}_{j \ell, j \ell}(R) +\epsilon_{j}+\ell(\ell+1)/(2\mu R^2)$ in Eq.~\ref{eq:Schrodinger_Eq} of the main text,  below the distance $R_{\textrm{sr}}=30$~a.u.. The parameter $D_{LJ}=0.15$~a.u. is chosen equal for all channels, leading to a sufficiently deep potential to minimize the probability density at this short range, and thus its influence on the eigenvalues. 

\begin{table}
\caption{Parameters of the Lennard-Jones potentials (Eq.~\ref{eq:LJ_pot}) extending below $R_{\textrm{sr}}=30$~a.u. the diagonal elements $W^{J}_{j \ell, j \ell}(R)$. The energies $E(\infty)$ correspond to the location of the $j=0,1,2$ of NaK. }
\begin{ruledtabular}
\begin{tabular}{l|l|l}
Diagonal term $W_{jl,j'l'}$ & $C_{LJ}$ (a.u.)& $E(\infty)$ a.u. \\ \hline
$W^1_{01,01}$ & 43685 & 0 \\
$W^1_{10,10}$ & 43809 & 8.702$\times 10^{-7}$ \\
$W^1_{12,12}$ & 46704 & 8.702$\times 10^{-7}$ \\
$W^1_{21,21}$ & 46949 & 2.610$\times 10^{-6}$\\
$W^1_{23,23}$ & 46940 & 2.610$\times 10^{-6}$\\
\end{tabular}
\end{ruledtabular}
\label{tab:short-range}
\end{table}


\section{Computed bound and quasibound levels of the $1^2A''$ state }
\label{app:levels_Adoubleprime}

We display in Table~\ref{tab:Adoubleprime} the energies $E_{n}^J$ (with respect to the NaK($b$, $v_b=0, j_b=1$)+K($4s$)) limit, the rotational constants $B_{n}^J$ multiplied by $10^{3}$ (Eq.~\ref{eq:rotconstant} of the main text), and the partial norms $C_{j \ell,n}^{J}$ (Eq.~\ref{eq:partialnorm} of the main text) of the weakly-bound vibrational levels (numbered from the uppermost one with a negative index, $n=-13$ to -1) of the $1^2A''$ ($J=1$) state, and located below the NaK($b$, $v_b=0, j_b=0$)+K($4s$) limit (at -0.19082~cm$^{-1}$ on this scale).  We see from the partial norms that the corresponding channel functions are more mixed than for the levels of $A'$ symmetry, due to the more significant anisotropy of the $A''$ PES (Fig.~\ref{fig:PES-cuts} of the main text) compared to the $A'$ one.

The same quantities are displayed for predissociating resonances ($n=$1 to 5) located between the NaK($b$, $v_b=0, j_b=0,1$)+K($4s$) limits.
\begin{table*}[h]
\caption{Energies $E_{n}^J$ (with respect to the NaK($b^3\Pi$, $v=0, j=1$)+K($4s$) limit), rotational constants $B_{n}^J$ multiplied by $10^{3}$ of the super dimer (Eq.~\ref{eq:rotconstant} of the main text), and partial norms $C_{j \ell,n}^{J}$ (Eq.~\ref{eq:partialnorm} of the main text) of weakly-bound vibrational levels (numbered from the uppermost one with a negative index, $n=-13$ to -1) of the $1^2A''$ ($J=1$) state, and located below the NaK($b^3\Pi$, $v=0, j=0$)+K($4s$) limit (at -0.19082~cm$^{-1}$ on this scale).  The stars label levels which could be seen in the proposed PA scheme. The same quantities are displayed for predissociating resonances ($n=$1 to 5) located between the NaK($b^3\Pi$, $v=0, j=0,1$)+K($4s$) limits }
\begin{ruledtabular}
\begin{tabular}{ >{\raggedright\arraybackslash}p{0.05\linewidth} | >{\raggedright\arraybackslash}p{0.1\linewidth} | >{\raggedright\arraybackslash}p{0.1\linewidth}| >{\raggedright\arraybackslash}p{0.1\linewidth}|
>{\raggedright\arraybackslash}p{0.1\linewidth}|
>{\raggedright\arraybackslash}p{0.1\linewidth}|
>{\raggedright\arraybackslash}p{0.1\linewidth}|     
>{\raggedright\arraybackslash}p{0.1\linewidth}}
$n$&$E_{n}^1$ (cm$^{-1}$)& $10^{3} \times B_{n}^1$(cm$^{-1})$ & $C_{01,n}^1$& $C_{10,n}^1$ & $C_{12,n}^1$& $C_{21,n}^1$& $C_{23,n}^1$ \\ \hline
-13&  -1.61635 & 4.4511 & 0.17597 & 0.00018 & 0.00014 & 0.74555 & 0.07816 \\
-12&  -1.58641 & 4.4278 & 0.05612 & 0.00035 & 0.00082 & 0.16056 & 0.78215 \\
-11&  -1.37263 & 3.8179 & 0.79210 & 0.00570 & 0.00662 & 0.07824 & 0.11734 \\
-10&  -1.22167& 3.7828 & 0.01156 & 0.53788 & 0.44944 & 0.00033 & 0.00079 \\
-9&  -0.92823 & 3.5848 & 0.00037 & 0.44651 & 0.55201 & 0.00044 & 0.00068 \\
-8&  -0.73189 & 2.9503 & 0.82896 & 0.00477 & 0.00544 & 0.08911 & 0.07172 \\
-7&  -0.69520 & 3.6467 & 0.18199 & 0.00238 & 0.00184 & 0.76137 & 0.05242 \\
-6&  -0.66942 & 3.6542 & 0.05394 & 0.00039 & 0.00102 & 0.11559 & 0.82906 \\
-5&  -0.57678 & 2.9302 & 0.01362 & 0.55454 & 0.43087 & 0.00072 & 0.00025 \\
-4& -0.39169& 2.7326 & 0.01100 & 0.42165 & 0.56589 & 0.00092 & 0.00054 \\
-3& -0.37529 & 2.1176 & 0.93572 & 0.00944 & 0.00708 & 0.01980 & 0.02796 \\
-2& -0.22538 & 1.2663 & 0.94730 & 0.02153 & 0.01475 & 0.00719 & 0.00923 \\
-1& -0.21034 & 2.1105 & 0.04245 & 0.56686 & 0.38888 & 0.00122 & 0.00059 \\ 
  \hline \hline
  1&  -0.11444 & 1.9803 & 0.01875 & 0.32995 & 0.52482 & 0.09089 & 0.03558 \\
  2&  -0.08536 & 2.7102 & 0.05273 & 0.00090 & 0.00197 & 0.20290 & 0.74150 \\
  3&  -0.04464 & 0.8147 & 0.40563 & 0.37807 & 0.21577 & 0.00032 & 0.00021 \\
  4&  -0.01224 & 0.8474 & 0.09115 & 0.33877 & 0.56988 & 0.00010 & 0.00009 \\
  5&  -0.00104 & 0.4051 & 0.00904 & 0.75482 & 0.23612 & 0.00001 & 0.00001
\end{tabular}
\end{ruledtabular}
\label{tab:Adoubleprime} 
\end{table*}

\section{Examples of radial wave functions of the K$\cdots$N\lowercase{a}K complex}
\label{app:radial-wf}

\begin{figure}
    \centering
    \includegraphics[scale=0.6]{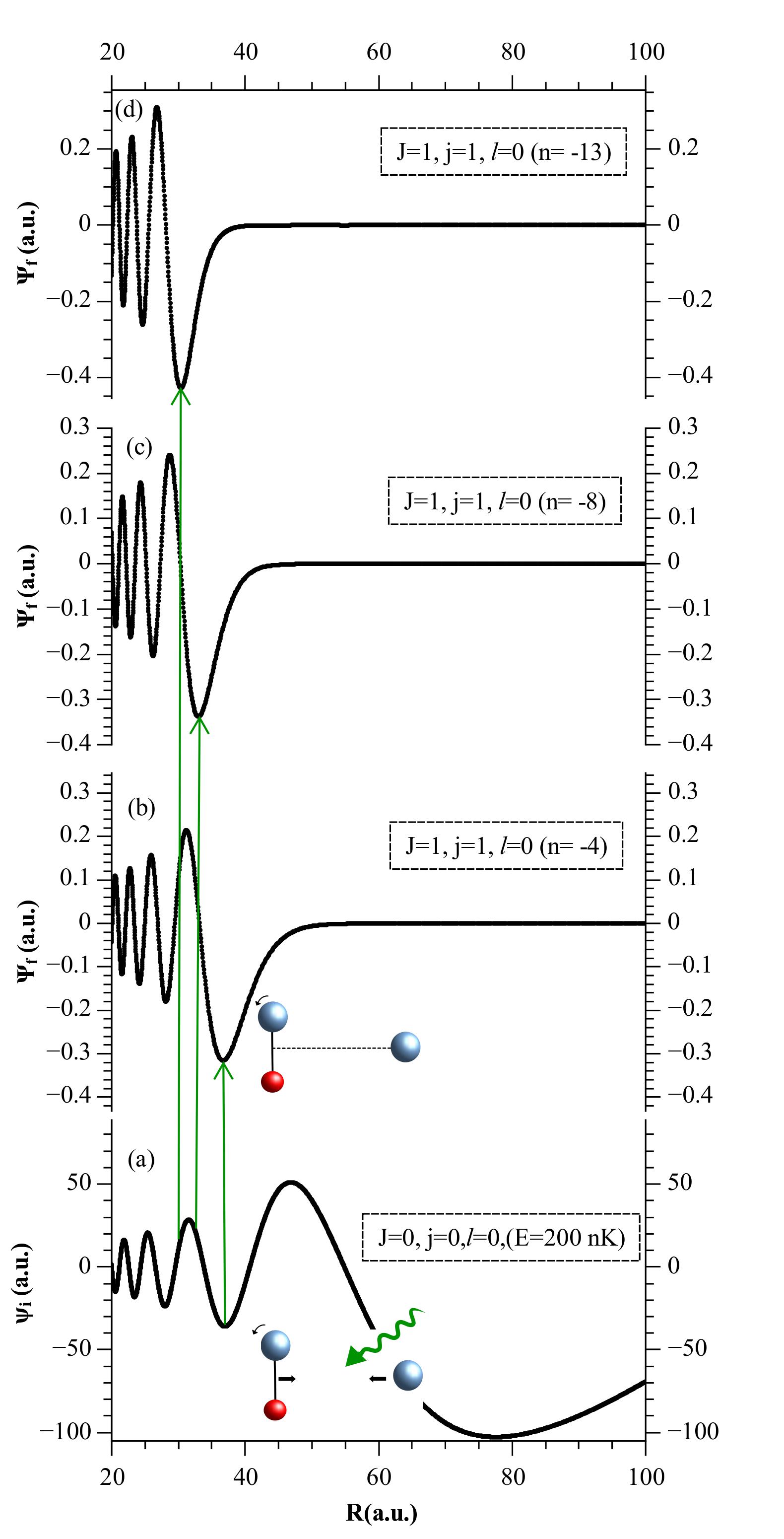}
    \caption{(a) Energy-normalized radial wave function for the initial scattering state of PA, associated to the NaK($X$, $v_X=0, j_X=0$)+K($4s$) threshold at a collision energy $E=k_B\times 200$~nK; (b)-(d) Radial wave functions of the final weakly-bound states of NaK $\cdots$K associated with the energy levels $n=-4,-8,-13$ in Table~\ref{tab:Aprime}. The vertical arrows illustrate the good matching of the first oscillation (starting from the right-hand side) of the bound-state wave functions with the continuum wave function.}
    \label{fig:WF}
\end{figure}

We display in Fig.~\ref{fig:WF} the radial wave functions for some of the weakly bound levels of NaK $\cdots$K of $A'$ symmetry corresponding to large computed PA rates, and on the same distance range, the energy-normalized radial wave function of the initial scattering state of PA with a collision energy $E=k_B\times 200$~nK. The spatial overlap between the initial state wave function and the final state wave functions of the proposed PA process is dominated by the value of the bound-state wave function at its first maximum from the right-hand side. For the chosen levels, a good overlap with the scattering wave function is indeed expected, as suggested by the vertical arrows, thus providing the main contribution to the computed PA rate. 

\section{Predissociating resonances}
\label{app:prediss-res}

\begin{figure}
    \centering
\includegraphics[scale=0.6]{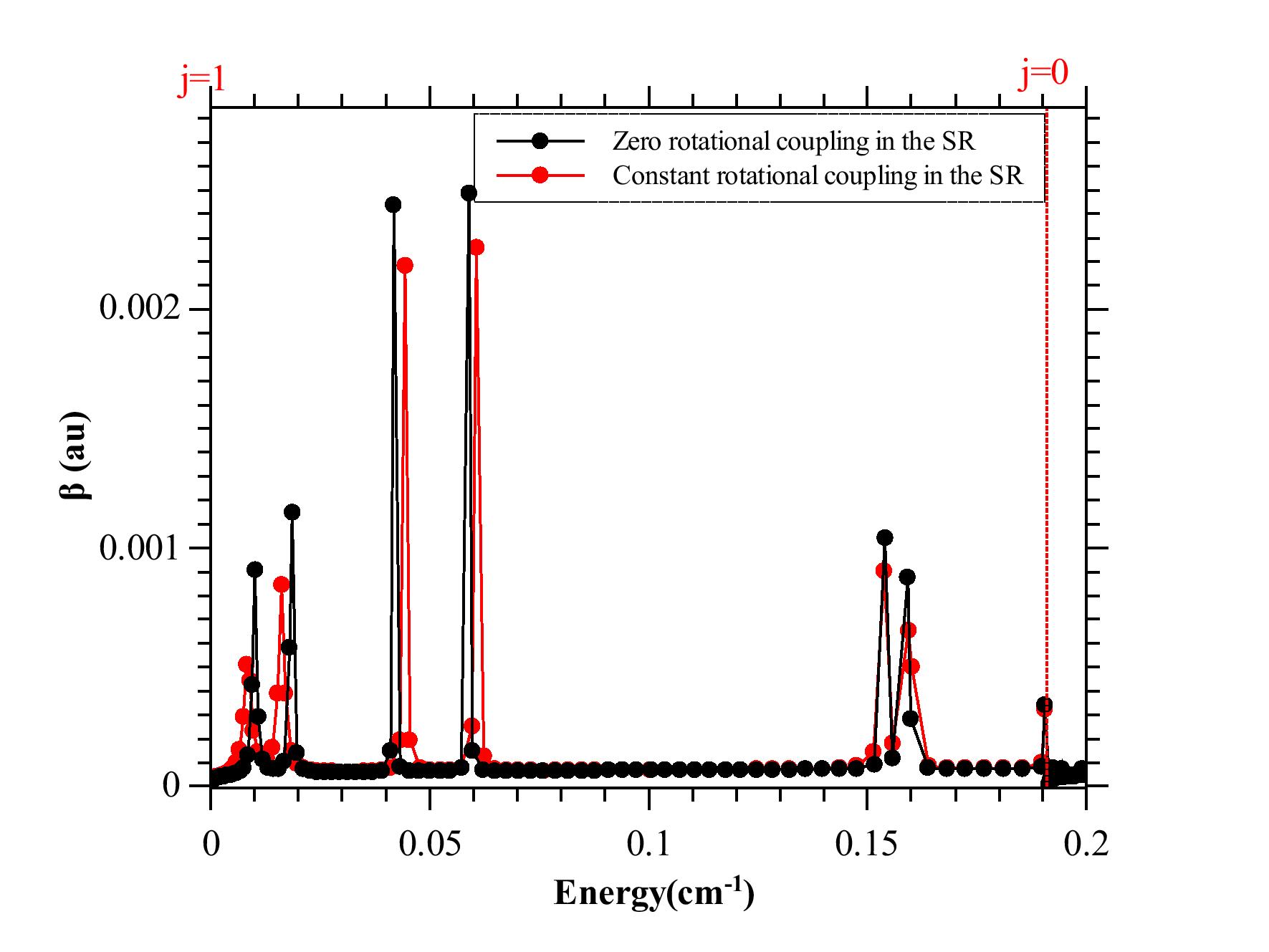}
    \caption{Eigenvalues above the lowest dissociation limit, resulting from the MFGH solution of the coupled equations, after application of the stabilization approach. The predissociation resonances appear as isolated peaks between the asymptotic channels $j_b=0,\ell=1$ and $j_b=1,\ell=0$ in the excited state $3^2A'$ correlated to NaK($b$,$v_b=0$)+K($4s$). All the other dots correspond to the discretization of the continuum induced by the finite grid in $R$ used in MFGH. Via the MFGH method, we assign to each eigenvalue a parameter $\beta$ (calculated from Eq.~\ref{eq:rotconstant} of the main text) analogous to the rotational constant for a physical level. We compare two cases for the off-diagonal terms of the potential energy matrix at short distances $R<R_{\textrm{sr}}$: vanishing couplings (black circles), and constant couplings (red circles).}
    \label{fig:Quasi-bound}
\end{figure}

We assigned predissociating resonances, \textit{i. e.} quasibound levels located between the NaK($b$, $v_b=0, j_b=0$)+K($4s$) and the NaK($b$, $v_b=0, j_b=1$)+K($4s$) limits, using the MFGH method and the stabilization method, as in Ref.~\onlinecite{osseni2009}. In brief, by extending the upper bound of the grid to sufficiently large distances $R_{\textrm{max}}=1000$~a.u. to properly describe the continuum with a large number of artificial levels, we identified stable eigenvalues by plotting the quantity $\beta$, identical to $B_n^J$ from Eq.~\ref{eq:rotconstant} of the main text, as a function of energy (Fig.~\ref{fig:Quasi-bound}). It is not a rotational constant in a rigorous way, as the related eigenfunctions involve a continuum part, but it provides a convenient way to estimate the width of the resonances via the energy range on which the continuum is perturbed by the relevant bound level. We see that the first four resonances are quite narrow ($\approx 0.001$~cm$^{-1}$), and could probably be detected. Note that in these calculations, we set the off-diagonal terms of the potential energy matrix to zero for $R<R_{\textrm{sr}}$. We see in the figure that the results are almost unchanged if these off-diagonal terms are kept constant in the short-range. 

\newpage
 \clearpage
 \section*{References}
	\bibliographystyle{unsrt}
	\bibliography{bibliocold,bibnote}

\begin{thebibliography}{10}
\bibitem{Mvalue}
As there is no external field in our model, the value of $M$ is arbitrary and
  is omitted in the forthcoming ket notations, for simplicity.
\end{thebibliography}

\end{document}